\documentclass[authoryear,preprint,12pt]{elsarticle}

\journal{Physical Communication}
\setcitestyle{numbers}
\usepackage{amsthm}
\usepackage{epsfig,amsfonts,subfigure,color}
\usepackage[nolist]{acronym}
\usepackage{graphicx,amssymb,amsmath,mathrsfs}
\usepackage{verbatim}
\usepackage{hyperref}
\usepackage{tikz,bm,color}
\usepackage{balance}
\usepackage{booktabs}
\usepackage{multirow}
\usepackage{siunitx}
\usepackage{paralist}
\usepackage{placeins}
\usepackage{algorithm}
\usepackage{algpseudocode}

\usepackage{pgfplots,relsize}
\usepgfplotslibrary{fillbetween}

\usetikzlibrary{intersections}

\newcommand{\cer}[1]{$\mathrm{CER}\approx 10^{#1}$}

\hypersetup{
	colorlinks=true,       
	linkcolor=black,          
	citecolor=black,        
	filecolor=magenta,      
	urlcolor=blue           
}

\newcommand{\snr}{E_b/N_0}
\newcommand{\ff}[1]{\mathbb{F}_{#1}}
\newcommand{\code}{\mathcal{C}}
\newcommand{\codelist}{\mathcal{L}}
\newcommand{\argmax}[1]{\underset{#1}{\mathrm{arg \, max}}\,}

\newcommand{\bx}{\mathbf{x}}
\newcommand{\by}{\mathbf{y}}
\newcommand{\bG}{\mathbf{G}}
\newcommand{\bu}{\mathbf{u}}
\newcommand{\dmin}{d_{\mathrm{min}}}
\newcommand{\CER}{\mathrm{CER}}

\newcommand{\dumx}{\mathtt{x}}
\newcommand{\Prob}{\mathbb{P}}
\newcommand{\mcc}{}

\begin{document}

\begin{acronym}
\acro{ARA}{accumulate-repeat-accumulate}
\acro{AR3A}{accumulate-repeat-3-accumulate}
\acro{ARJA}{accumulate-repeat-jagged-accumulate}
\acro{AWGN}{additive white Gaussian noise}
\acro{B-DMC}{binary-input discrete memoryless channels}
\acro{B-MC}{binary-input memoryless channels}
\acro{BCJR}{Bahl-Cocke-Jelinek-Raviv}
\acro{BEAST}{a bidirectional efficient algorithm for searching trees}
\acro{BP}{belief propagation}
\acro{BPSK}{binary phase shift keying}
\acro{BCH}{Bose-Chaudhuri-Hocquenghem}
\acro{bi-AWGN}{binary-input additive white Gaussian noise}
\acro{CC}{convolutional code}
\acro{CRC}{cyclic redundancy check}
\acro{CCSDS}{Consultative Committee for Space Data Systems}
\acro{CER}{codeword error rate}
\acro{DE}{density evolution}
\acro{EXIT}{extrinsic information transfer}
\acro{eMBB}{enhanced Mobile Broadband}
\acro{H-ARQ}{hybrid automatic repeat request}
\acro{FFT}{fast Fourier transform}
\acro{GA}{Gaussian approximation}
\acro{LDPC}{low-density parity-check}
\acro{LLR}{log-likelihood ratio}
\acro{LTE}{long term evolution}
\acro{LVA}{list Viterbi algorithm}
\acro{ML}{maximum likelihood}
\acro{NR}{New Radio}
\acro{NA}{normal approximation}
\acro{OSD}{ordered statistics decoding}
\acro{PEG}{progressive edge growth}
\acro{RA}{random access}
\acro{RCB}{random coding bound}
\acro{SC}{successive cancellation}
\acro{SCL}{\ac{SC} list}
\acro{SNR}{signal-to-noise ratio}
\acro{SISO}{soft-input soft-output}
\acro{SPB}{sphere packing bound}
\acro{TB}{tailbiting}
\acro{WAVA}{wrap-around Viterbi algorithm}
\acro{SE}{spectral efficiency}
\acro{BRGC}{binary reflected gray code}
\acro{PAS}{probabilistic amplitude shaping}
\acro{CCDM}{constant composition distribution matching}
\acro{SMDM}{shell mapping distribution matching}
\acro{DM}{distribution matcher}
\acro{RCU}{random coding union}
\acro{MC}{metaconverse}
\acro{NA}{normal approximation}
\end{acronym}

\begin{frontmatter}

\title{Efficient Error-Correcting Codes in the Short Blocklength Regime}

\author[DLR,TUM]{Mustafa Cemil Co\c{s}kun}
\author[Chalmers]{Giuseppe Durisi}
\author[DLR]{Thomas Jerkovits}
\author[DLR]{Gianluigi Liva}
\author[ZETA]{William Ryan}
\author[ZETA]{Brian Stein}
\author[TUM]{Fabian Steiner}

\address[DLR]{Institute of Communications and
	Navigation of the German Aerospace Center (DLR), M\"unchner Strasse 20, 82234 We{\ss}ling, Germany. 	Email: \texttt{\{mustafa.coskun,thomas.jerkovits,gianluigi.liva\}@dlr.de}.}
\address[Chalmers]{Department of Electrical Engineering, Chalmers University of Technology, Gothenburg, Sweden. Email: \texttt{durisi@chalmers.se}.}
\address[TUM]{Institute of Communication Engineering of the Technical University of Munich (TUM), Theresienstrasse 90, 80333 Munich, Germany. Email:\texttt{fabian.steiner@tum.de}.}
\address[ZETA]{Zeta Associates Inc., 10302 Eaton Place, Suite 500 Fairfax, VA 22030.}

\begin{abstract}
The design of block codes for short information blocks (e.g., a thousand or less information bits) is an open research problem that is gaining relevance thanks to emerging applications in wireless communication networks. In this paper, we review some of the most promising code constructions targeting the short block regime, and we compare them with both finite-length performance bounds and classical error-correction coding schemes. The work addresses  the use of both binary and high-order modulations over the additive white Gaussian noise channel. We will illustrate how to effectively  approach the theoretical bounds with various performance versus decoding complexity tradeoffs.
\end{abstract}

\begin{keyword}
	Short packets \sep error-correcting codes \sep finite-length performance bounds \sep coded modulation.
\end{keyword}

\end{frontmatter}
\section{Introduction}\label{sec:intro}

During the past sixty years, a formidable effort  has been focused on the research of capacity-approaching error correcting codes \cite{Shannon48}.
Initially, the attention was directed to short and medium-length linear block codes \cite{berlekamp74:key} (with some notable exceptions, see, e.g., \cite{Elias54:EFC,Gallager63:LDPC}), mainly for complexity reasons.
As the idea of code concatenation \cite{Forney66:Concatenated} became established in the coding community \cite{Costello07:PROC}, the design of long channel codes became a viable solution to approach the channel capacity. The effort resulted in a number of practical code constructions allowing reliable transmission at fractions of a decibel from the Shannon limit \cite{Berrou93:TC,MacKay99:LDPC,Richardson01:Design,richardson01:capacity,Luby01:LDPC,Pfister05:capacity,Pfister07:ARA,Arikan09:Polar,Lentmaier10:CLDPC,Kudekar11:CLDPC} with low-complexity (sub-optimum) decoding.

The interest in short and medium blocklength codes (i.e., codes with dimension $k$ in the range of $50$ to $1000$ bits) has been rising again recently, mainly due to emerging applications that require the transmission of short data units. Examples of such applications are machine-type communications, smartmetering networks, the \mcc{Internet} of things, remote command links and messaging services (see, e.g., \cite{Decola11:CCSDS,Boccardi14:MAG,Paolini15:MAG,Durisi16:Short}).
Due to these new emerging applications, renewed interest has been placed not only in the design of efficient short codes, but also in the development of tight bounds on the performance attainable by the best error correcting schemes, for a given blocklength and rate \cite{Dolinar98:BOUNDS,Duman98,SasonShamai06:BOUNDS,Polyanskiy10:BOUNDS}. Tight bounds are now available as benchmarks not only for the unfaded \ac{AWGN} case but also for fading channels \cite{Durisi14,Durisi16}.

When the design of short iteratively-decodable codes is attempted, it  turns out that some classical code construction tools that have been developed for turbo-like codes tend to fail to provide codes with acceptable performance. This is the case, for instance, for density evolution \cite{Richardson08:BOOK} and \ac{EXIT} charts \cite{tenBrink01:EXIT}, which are well-established techniques to design powerful long \ac{LDPC} and turbo codes. The reason is the asymptotic (in blocklength) nature of density evolution and \ac{EXIT} analysis, which fail to  model accurately the iterative decoder in the short blocklength regime. However, competitive \ac{LDPC} and turbo code designs for moderate length and small blocks have been proposed, mostly based on heuristic construction techniques \cite{Sadjadpour00:shortTC,Koetter03:Wheel,Ryan04:eIRA,Liva05:Tanner_Milcom,Divsalar07:ShortProto,Liva08:QC_GLDPC,Bocharova09:shortLDPC,Divsalar14:ProtoRaptor,Jerkovits16:Short,Davey98:NonBinary,Berkmann00:Diss,Berkmann02:dualtrellis,Poulliat08:BinImag,Venkiah08:RPEG,Chen09:Hamilton,Costantini10:NBLDPC,kasai11:multiplicatively,Liva12:IRAq,Divsalar12:NonBinaryShort,Liva13:ShortTC,Matuz13:LRNBTC,Dolecek14:NBLDPCTIT}.
While iterative codes retain a large appeal due to their low decoding complexity, more sophisticated decoding algorithms \cite{Fossorier95:OSD,Fossorier04:BOX,Huber07:Multibase,Huber10:Multibase,Wu07:MRB,Liva14:OptProd,Tal15:ListPolar} are feasible for short blocks leading to solutions that are competitive with (if not superior to) iterative decoding of short turbo and \ac{LDPC} codes.

In this paper, we review some fundamental results on the performance achievable by codes in the short blocklength regime. This will allow us to lay the ground for a proper performance comparison among various codes and decoding algorithms. The comparison will be provided for the (unfaded) \ac{AWGN} channel case with both binary modulation and high-order modulations. In the former case, the goal is to compare pure code performance, whereas in the latter case we shall see how different coding schemes can be efficiently coupled with high-order modulations, with and without shaping. The performance comparison will be provided in terms of  block error rate, also referred to as \ac{CER}, versus \ac{SNR} with \ac{SNR} given either by the $E_b/N_0$ ratio (here, $E_b$ is the energy per information bit and $N_0$ the single-sided noise power spectral density) or by the $E_s/N_0$ ratio (with $E_s$ being the energy per modulation symbol).

The remaining part of the paper is organized as follows. Section \ref{sec:limits} reviews the fundamental limits for channel coding in the short blocklength regime. Some powerful classical short codes as well as efficient decoding algorithms are discussed in Section \ref{sec:codesclassic}. Modern code constructions tailored to the transmission of short blocks are presented in Section \ref{sec:codesmodern}. A comparison of various schemes is provided in Section \ref{sec:comparison}. Conclusions follow in Section \ref{sec:conclusions}.
\section{Finite-Length Performance Limits}\label{sec:limits}
In the following sections of the paper, with the exception of Section~\ref{sec:coded_modulation}, we will focus on the problem of how to optimally transmit $k$ bits of information using the discrete-time memoryless \ac{bi-AWGN} channel ~$n$ times
\begin{equation}\label{eq:io-relation}
  y_\ell=\sqrt{\rho} x_\ell+w_\ell, \quad \ell=1,\dots, n.
\end{equation}
Here, $\{w_\ell\}_{\ell=1}^n$, denotes a sequence of independent and identically distributed samples of the AWGN process. We shall assume that these samples are real-valued Gaussian random variables with zero mean and unit variance.
Each of the input symbols $\{x_\ell\}_{\ell=1}^n$ belongs to the binary set $\{-1,1\}$.
The constant $\rho$ models the transmit power and, hence, the SNR, since the noise has unit variance.
Finally, $\{y_\ell\}_{\ell=1}^n$ is the sequence of received symbols.

To convey the $k$ information bits, we use a $(n,k)$ coding scheme, which consists of:
\begin{inparaenum}[i)]
  \item An encoder that maps the $k$-bit message $J\in\{1,\dots,2^k\}$ into one out of $2^k$ $n$-dimensional codewords with elements in $\{-1,1\}$. 
  We shall refer to the set of codewords together  with the encoder as an $(n,k)$ code and to $n$ as the blocklength of the code.
  \item A decoder that maps the $n$ received symbols corresponding to the transmitted message~$J$ into an estimated $k$-bit message~$\widehat{J}$.
\end{inparaenum}

The message (codeword) error probability of a given $(n,k)$-coding scheme, which we denote by $\epsilon$, is
\begin{equation}\label{eq:CER}
  \epsilon= \Prob[\widehat{J}\neq J].
\end{equation}
We stress that different decoders may be applied to a given $(n,k)$ code, yielding different error probabilities.

The rate $R$ of an $(n,k)$ code is $R=k/n$. 
We also let $\epsilon^*(R,n)$ be the minimum error probability for which one can find a coding scheme with blocklength $n$ and rate~$R$.
This quantity describes the fundamental tradeoff between blocklength $n$, rate $R$, and error probability $\epsilon$ in the transmission of information.
Unfortunately, determining $\epsilon^*(R,n)$ exactly is a daunting task. 
Indeed, computing $\epsilon^*(R,n)$ for the bi-AWGN channel~\eqref{eq:io-relation} involves an exhaustive search over $\binom{2^n}{2^{nR}}$ codes, which is infeasible for values of $R$ and $n$ of practical interest.

However, the asymptotic behavior of $\epsilon^*(R,n)$ in the limit $n\rightarrow\infty$ for fixed $R$ is well understood---a result known as Shannon's coding theorem~\cite{Shannon48}.
Specifically, $\epsilon^*(R,n)$ vanishes in the limit $n\rightarrow\infty$ for all rates $R$ below the so-called channel capacity $C$~\cite{Shannon48}, whereas $\epsilon^*(R,n)\rightarrow 1$ as $n\rightarrow \infty$ for all rates $R$ above $C$.
In other words, the sequence of functions $f_n(R)=\epsilon^*(R,n)$ converges to a step function centered at $C$ in the limit $n\rightarrow \infty$.

For the bi-AWGN channel~\eqref{eq:io-relation}, the capacity $C$ (measured throughout the paper in bits per channel use) is given by
\begin{equation}\label{eq:capacity_bi_awgn}
  C=\frac{1}{\sqrt{2\pi}}\int e^{-z^2/2} \Bigl(1-\log_2\bigl(1+e^{-2\rho+2z\sqrt{\rho}}\bigr) \Bigr)\, 
  \mathrm{d}z.
\end{equation}
The achievability part of Shannon's coding theorem relies on a random coding argument and does not suggest practical capacity-approaching coding schemes.
However, several advances in the coding community over the last sixty years have resulted in several low-complexity coding schemes that approach capacity (see e.g.,~\cite{Richardson08:BOOK,RyanLin09:BOOK}).

In this paper, we are concerned with the less studied problem of how to approach $\epsilon^*(R,n)$ when the blocklength $n$ is short. 
For the problem to be well posed, we need ways to  estimate $\epsilon^*(R,n)$ accurately.
Characterizing $\epsilon^*(R,n)$ for a fixed blocklength $n$ is a classic problem in information theory, and many nonasymptotic upper (achievability) bounds and lower (converse) bounds are available for the bi-AWGN channel~\eqref{eq:io-relation}, such as Gallager's 
\ac{RCB}~\cite{gallager65-01a}, and Shannon's sphere-packing bounds `59 (SPB59)~\cite{Shannon59:SPB} and `67 (SPB67)~\cite{shannon67-a,Fossorier04:BOUNDS}.
Also, many nonasymptotic results are available on the error probability achievable using linear block codes and \ac{ML} decoding (see~\cite{SasonShamai06:BOUNDS} and reference therein).

Over the last ten years, a renewed interest in the performance of communication systems operating in the short-blocklength regime, has resulted in a significant improvement in the tightness of the best available achievability and converse bounds for many communication channels of practical interest, including the well-studied bi-AWGN channel~\eqref{eq:io-relation}.

To showcase such improvements, we will focus in this paper on two specific classes of bounds, namely converse bounds based on the so called 
\ac{MC} theorem~\cite[Thm.~26]{Polyanskiy10:BOUNDS},  and achievability bounds based on the \ac{RCU} bound~\cite[Thm.~16]{Polyanskiy10:BOUNDS}.

The \ac{MC} theorem provides a general framework that allows one to recover all previously known converse bounds on $\epsilon^*(R,n)$ (hence, its name).
The theorem exploits the existence of a fundamental relation between the problem of determining the error probability of a given code under \ac{ML} decoding and  binary-hypothesis testing~\cite{vazquez-vilar16-05a}. 
The resulting converse bound is parametric in an auxiliary output distribution (i.e., a marginal distribution on the output vector $[y_1,\dots,y_n]$), which, if chosen suitably, results in a remarkably tight bound that admits an efficient numerical implementation by using the saddlepoint approximation~\cite{vazquez-vilar18-07a}.

Similar to Gallager's \ac{RCB}, the \ac{RCU} bound relies on the analysis of the performance of a random coding ensemble under \ac{ML} decoding. 
As indicated by its name, a crucial step to obtain the \ac{RCU} is a judicious use of the union bound. 
An attractive feature of this bound, is that it generalizes naturally to mismatched decoding metrics~
\cite{martinez11-02a,scarlett14-05a}, which enables its use in practically relevant scenarios, such as pilot-assisted transmission over fading channels~\cite{Ostman2018:NC}.
Similarly to the \ac{MC} bound, the use of a saddlepoint approximation allows one to evaluate numerically the \ac{RCU} bound for the bi-AWGN channel~\eqref{eq:io-relation} in an efficient way~\cite{font-segura18-03a}. 

By characterizing both \ac{MC} and \ac{RCU} bounds in the asymptotic limit $n\rightarrow \infty$ for a fixed $R$, one obtains a more precise characterization of the behavior of $\epsilon^*(R,n)$ for large $n$ than the step-function approximation obtained via Shannon's capacity.
Specifically, one can show that for the bi-AWGN channel~\eqref{eq:io-relation},
\begin{equation}\label{eq:normal_approx}
  \epsilon^*(R,n)=Q\biggl(\frac{n(C-R)+(1/2)\log_2(n)+O(1)}{\sqrt{nV}}\biggr)
\end{equation}
where $C$ is the channel capacity given in~\eqref{eq:io-relation}, $V$ is the so-called channel dispersion
\begin{equation}\label{eq:channel_dispersion}
  V=\frac{1}{\sqrt{2\pi}}\int e^{-z^2/2} \Bigl(1-\log_2\bigl(1+e^{-2\rho+2z\sqrt{\rho}}\bigr)-C\Bigr)^2\mathrm{d}z
\end{equation}
$Q(\cdot)$ is the Gaussian $Q$ function and $O(1)$ comprises terms that can be upper-bounded by a constant for all sufficiently large $n$.
The approximation on $\epsilon^*(R,n)$ obtained by neglecting the $O(1)$ term in~\eqref{eq:normal_approx} is usually referred to as \ac{NA}.

\begin{figure*}[t]
	\begin{center}
		\includegraphics[width=\textwidth]{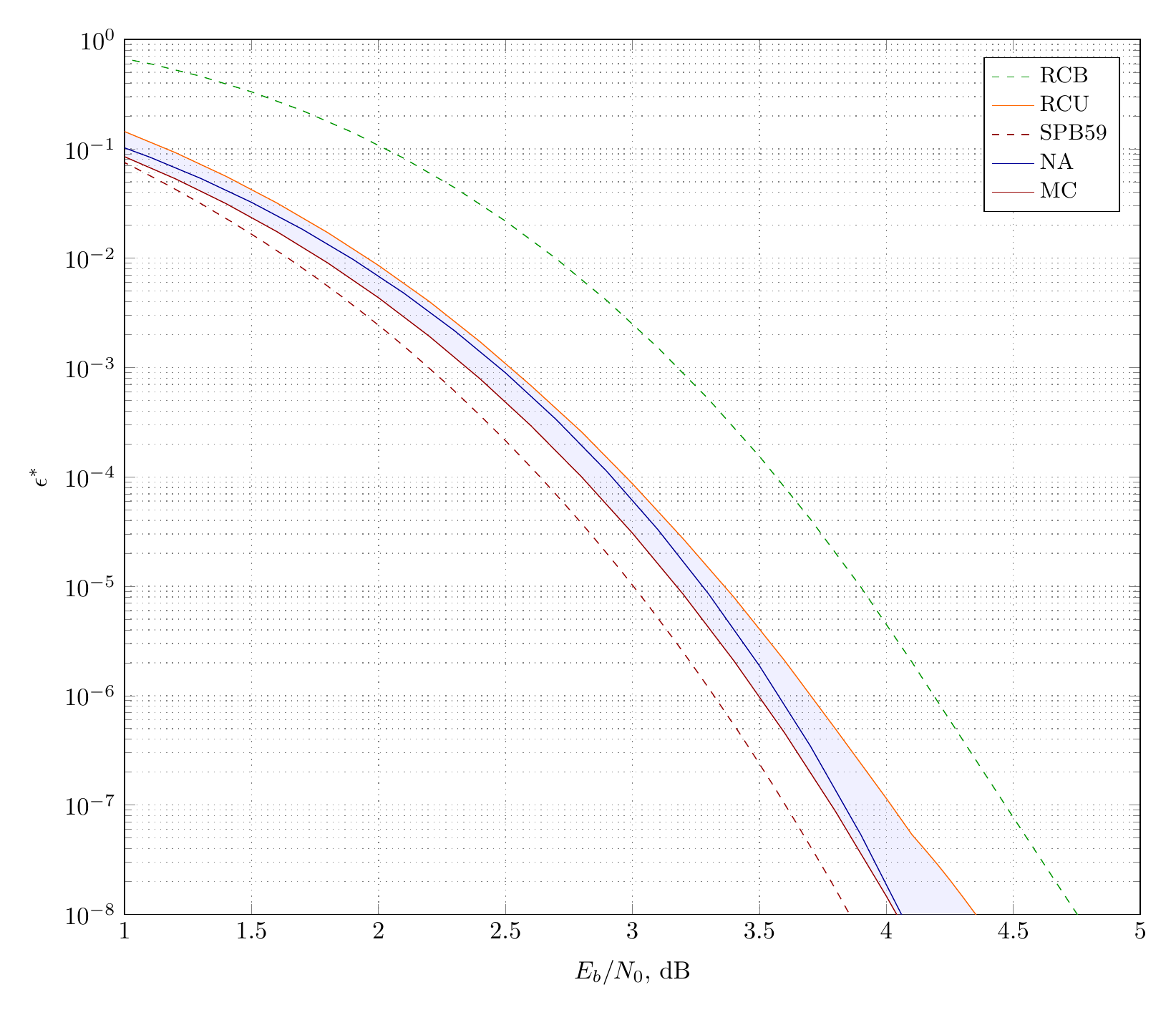}
	\end{center}
	\caption{Bounds on the minimum codeword error probability $\epsilon^*(R,n)$ vs. $E_b/N_0$ for  the case $n=128$ and $k=64$.}\label{fig:bounds_64128}
\end{figure*}

We now illustrate the tightness of the \ac{MC} bound, of the \ac{RCU} bound, and of the \ac{NA} through some numerical examples.
In Fig.~\ref{fig:bounds_64128}, we plot the \ac{MC} computed using the exponent-achieving auxiliary output distribution given in~\cite[Eq. (28)]{vazquez-vilar18-07a}, the \ac{RCU}, and the \ac{NA} for the case $n=128$ and $k=64$ (hence, $R=1/2$).
Here, and throughout the paper, the bounds on $\epsilon^*(R,n)$ given by the \ac{MC} and the \ac{RCU} are plotted as a function of the energy per information bit 
\begin{equation}\label{eq:energy_per_bit_def}
  \frac{E_b}{N_0}=\frac{\rho}{2R}.
\end{equation}
For comparison we also illustrate the SPB59 (which, for the parameters chosen in the figure, is tighter than the adaptation of the SPB67 bound given in~\cite{Fossorier04:BOUNDS}) and Gallager's RCB.
As one sees from the figure, the \ac{MC} and the \ac{RCU} bounds delimit tightly the $\epsilon^*(R,n)$ that is achievable for the chosen blocklength and information payload for a large range of $E_b/N_0$ values. 
For example, the two bounds predict that the minimum energy per bit to operate at a CER of $10^{-6}$ is between $3.5$ dB and $3.7$ dB. 
The SPB59 and the RCB are looser and give the wider range $3.3$ dB and $4.2$ dB.
Note also that the NA provides an accurate estimate of the minimum codeword error probability, which lies between the \ac{MC} and the \ac{RCU} bounds. 
As we shall see, this is not a general phenomenon and one can find practically relevant scenarios for which the \ac{NA} ceases to be as accurate.

\begin{figure}[h]
  \centering
    \includegraphics[width=.9\textwidth]{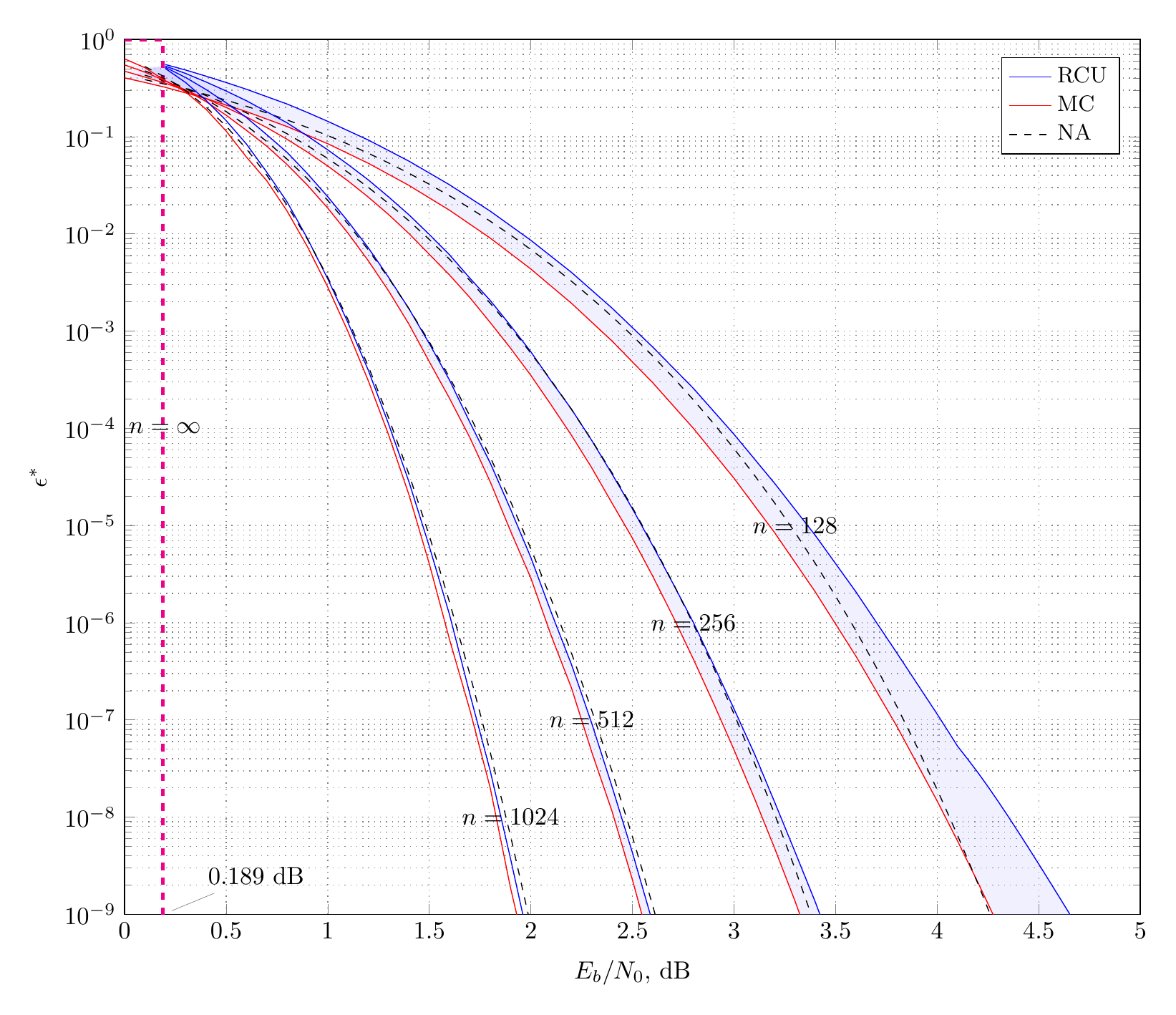}
  \caption{Bounds on the codeword error probability vs. $E_b/N_0$ for the case $R=1/2$ and $n\in \{128,256,512,1024\}$. Note that  $E_b/N_0=0.189$ dB is the minimum {$E_b/N_0$ required} to transmit reliably at rate $R=1/2$ in the limit $n\rightarrow\infty$.}
  \label{fig:charts_gd_bi-awgn-conv_na_alt}
\end{figure}

In Figure~\ref{fig:charts_gd_bi-awgn-conv_na_alt}, we plot the \ac{MC}, the \ac{RCU}, and the \ac{NA} for $R=1/2$ and $n\in\{128,256,512,1024\}$. 
As the blocklength increases, the gap between \ac{MC} and \ac{RCU} diminishes. 
Also Figure~\ref{fig:charts_gd_bi-awgn-conv_na_alt} allows one to estimate the speed at which $\epsilon^*(R,n)$ converges to a step function centered at $E_b/N_0=0.189$ dB, which is the minimum {$E_b/N_0$ required} to communicate at a rate $C=1/2$ in the asymptotic limit $n\rightarrow\infty$.
The gap to the asymptotic limit for $n=1024$ is about $1.4$ dB at a CER of $10^{-6}$.
The \ac{NA} is accurate in all the scenarios considered in the figure.

\begin{figure}[h]
  \centering  
    \includegraphics[width=.9\textwidth]{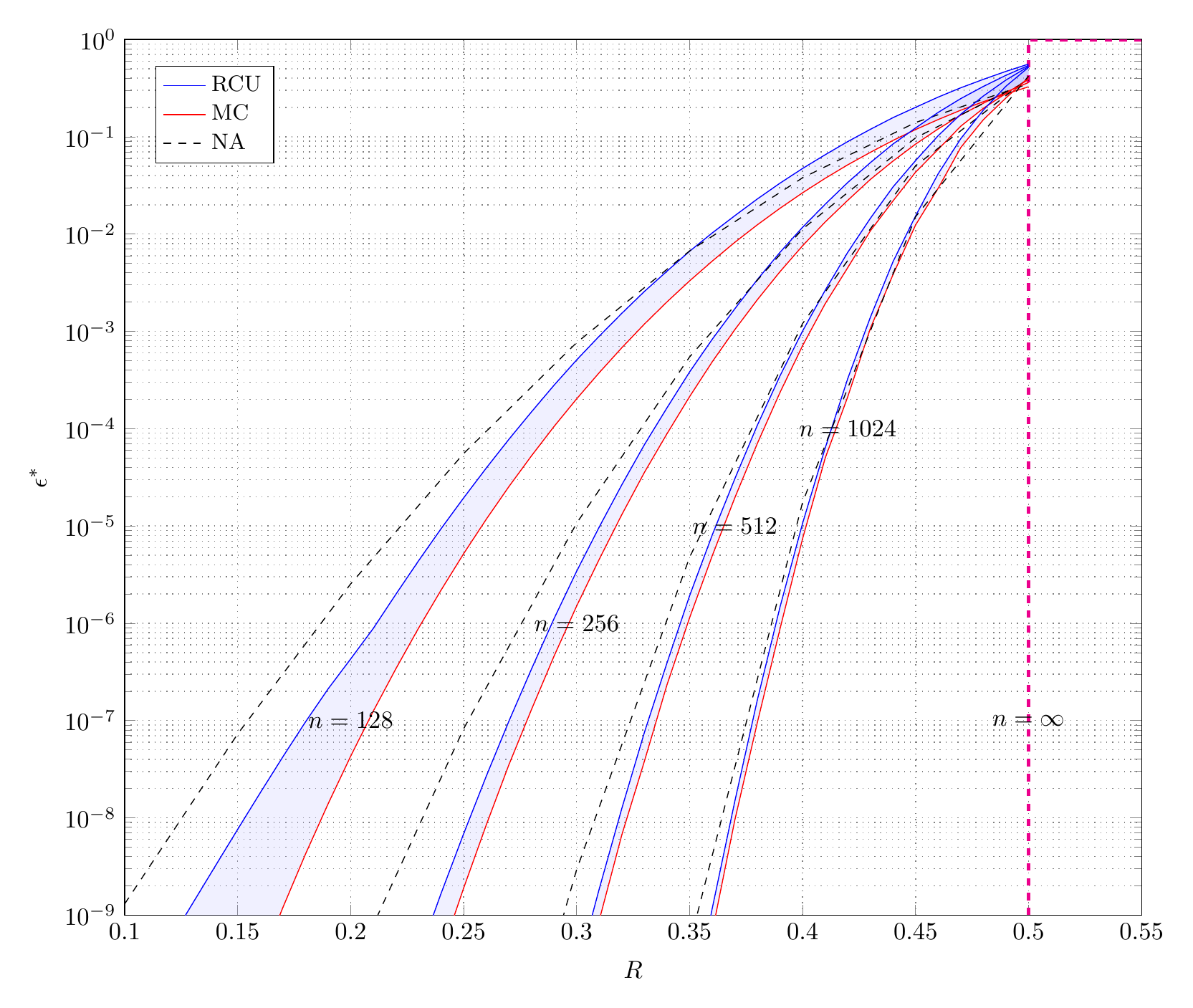}
  \caption{Bounds on the codeword error probability vs. rate for the case $\rho=0.189$ dB and $n\in \{128,256,512,1024\}$. Note that  $R=1/2$ is the channel capacity.}
  \label{fig:charts_gd_rate_vs_epsilon_na}
\end{figure}

Finally, in Figure~\ref{fig:charts_gd_rate_vs_epsilon_na} we plot $\epsilon^*(R,n)$ as a function of the rate $R$ for a fixed SNR $\rho=0.189$ dB, i.e., the SNR value for which capacity is $1/2$, and for $n\in\{128,256,512,1024\}$.
As in the previous figure, the bounds become increasingly tight as $n$ grows. 
One can also see that the \ac{NA} loses accuracy  when one operates at small 
error probability and small $R$---a relevant scenario for ultra-reliable low-latency communications.
 
\section{Classical Short Codes}\label{sec:codesclassic} 

In this section, we will review a few approaches for efficient transmission at short blocklengths which rely on (or can be applied to) classical error-correcting code families. The first approach is based on a general decoding algorithm called \ac{OSD} \cite{Fossorier95:OSD} that can be applied to any (binary) linear block code. As we shall see, \ac{OSD} delivers near-\ac{ML} performance for short codes with manageable complexity. 
However, \ac{OSD} becomes unfeasible when the blocklength increases. The second approach relies on \ac{TB}  \acp{CC} and an efficient near-\ac{ML} decoding algorithm based on the recursive application of Viterbi decoding to the \ac{TB} trellis of the code.

\subsection{Short Algebraic Codes under Ordered Statistics Decoding}\label{sec:codesclassic:OSD} 
Consider an $(n,k)$ binary linear block code $\code$. Under \ac{ML} decoding, the decision is given by
\begin{equation} \label{eq:short_codes:ML_rule}
\hat{\bx}=\argmax{\bx\in\code}p\left(\by|\bx\right) 
\end{equation}
with $p\left(\by|\bx\right)=\prod_{\ell=1}^{n}p(y_\ell|x_\ell)$ being the channel transition probability (we assume the channel to be an arbitrary binary-input memoryless channel).
The evaluation of \eqref{eq:short_codes:ML_rule} involves a number of computations that grow exponentially in $k$, unless the code exhibits some structure that enables an efficient implementation of the \ac{ML} search.
\ac{OSD} reduces the decoding complexity by limiting the search to a subset of the codewords, i.e., to a list $\codelist \subset \code$. 
Hence, decoding reduces to 
\begin{equation} \label{eq:short_codes:list_rule}
\hat{\bx}=\argmax{\bx\in\codelist}p\left(\by|\bx\right). 
\end{equation}
The decoding complexity is directly related to the list size. \ac{OSD} uses a particularly effective approach for the list construction, which is based on ranking the symbol-wise channel observations in decreasing order of reliability \cite{Fossorier95:OSD,LinCostello04}. The received vector $\by$ is permuted accordingly, yielding a vector $\by'$ whose first $k$ components are the most reliable channel observations. The columns of the code generator matrix $\bG$ are permuted accordingly. The permuted generator matrix $\bG'$ is then put in systematic form.\footnote{This may require additional column permutations, which shall be applied to  $\by'$ too; this step is required if the $k$ leftmost columns of $\bG'$ are not linearly independent. The additional permutations aim at having in the first $k$ positions of  $\by'$ the most reliable information set.} The first $k$ observations in $\by'$ are used to obtain (via bit-by-bit hard detection) a $k$ bit vector $\bu'$. All error patterns of Hamming weight up to $t$ (where $t$ is a parameter of the \ac{OSD} algorithm) are then added to $\bu'$, generating a set of vectors of cardinality $\sum_{i=0}^t {k \choose i}$. Each vector is then encoded via the systematic form of  $\bG'$, yielding the list $\codelist$. 
Typically, the \ac{OSD} parameter $t$ is kept small because the list size grows quickly with $t$.
\ac{OSD} relies on the idea that, if one  takes a hard decision on the most reliable channel observations, only few errors are typically observed, whereas the majority of the errors introduced by the channel are typically  associated with the least reliable channel outputs.

\begin{figure}[h]
	\begin{center}
		\includegraphics[width=0.9\textwidth]{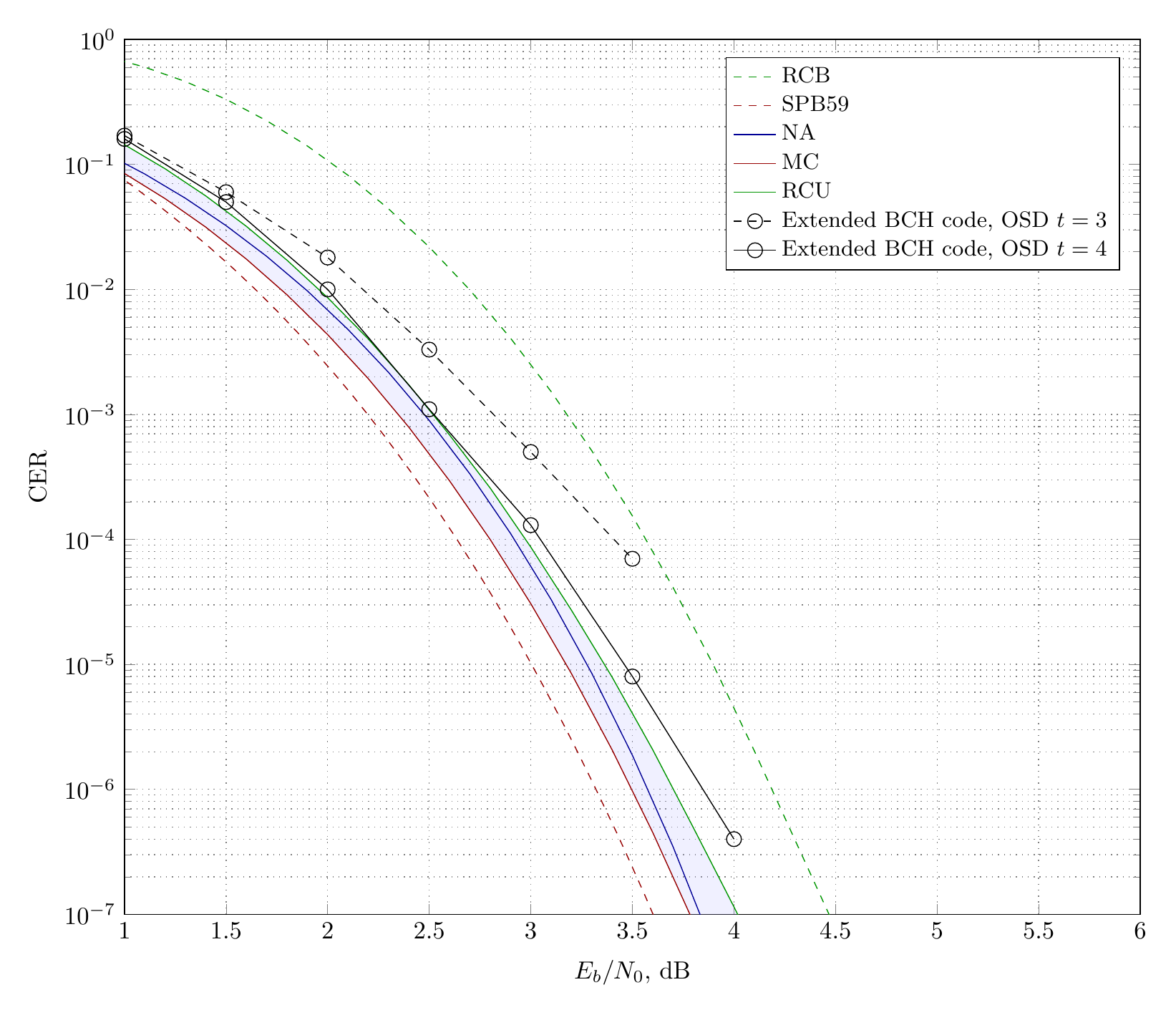}
	\end{center}
	\caption{Codeword error rate vs. $\snr$ for $(128,64)$ extended \ac{BCH} code under \ac{OSD} with $t=3$ and $t=4$, \ac{bi-AWGN} channel.}\label{fig:64128_BCH}
\end{figure}

\ac{OSD} works remarkably well with short codes, enabling near-\ac{ML} decoding for small values of the parameter $t$. However, as the blocklength grows, $t$ must be increased to keep the decoder performance close to optimal. For example, while for the $(24,12)$ Golay code choosing $t=2$ is enough to approach the \ac{ML} decoding limit, for a $(128,64)$ extended \ac{BCH} code one needs to set $t$ as large as $4$. Figure \ref{fig:64128_BCH} shows the performance in terms of \ac{CER} vs. $\snr$ for a $(128,64)$ extended \ac{BCH} code with $t=3$ and $t=4$ on the \ac{bi-AWGN} channel. For the case of $t=4$, the performance is within $0.1$ dB from the \ac{NA} at $\CER\approx10^{-4}$. With $t=3$ the gap increases to $\approx 0.5$ dB.

\ac{OSD} does not require any specific code property (besides linearity). However, some knowledge of the code distance spectrum can be used to simplify the decoder by introducing an early stopping criterion. Consider the example of the transmission over the \ac{bi-AWGN} channel. 
Assuming that the coded bits are mapped to symbols in the set $\{-1,+1\}$, the minimum Euclidean distance between modulated codewords is $2\sqrt{\dmin}$ where $\dmin$ is the code minimum Hamming distance. It follows that the list construction can be halted if a codeword at a Euclidean distance less than $\delta=\sqrt{\dmin}$ from the channel observation is generated. This simple trick yields remarkable savings on the average list size at moderate-large \acp{SNR}  \cite{Fossorier95:OSD}. 
Another simple approach to limit the complexity of \ac{OSD} consists of applying \ac{OSD} only if decoding with a lower-complexity algorithm has failed. The idea was explored, for instance, in \cite{Fossorier01:OSDLDPC,Baldi14:OSD} in the context of iterative decoding of \ac{LDPC} codes. Here, the 
\ac{OSD} can either intervene if the \ac{BP} decoder fails to converge to a valid codeword, or it can be even integrated within the iterative decoding algorithm by exploiting updated reliability estimates computed by the \ac{BP} decoder. A number of additional improvements on the efficiency of \ac{OSD} algorithms were further proposed during the past two decades (see, e.g., \cite{Fossorier04:BOX,Wu07:MRB} and the references therein).

\subsection{Tailbiting Convolutional Codes}

Short codes based on (compact) \ac{TB} trellises have been the subject of thorough studies from both a theoretical and a practical viewpoint \cite{Lin1998,Calderbank1999,Stahl99,Bocharova02,FCC15}. In particular, in \cite{Stahl99,Bocharova02} \ac{TB} \acp{CC} with excellent distance properties were proposed for various encoder memories, code rates, and blocklengths. The \ac{TB} structure of the trellis hinders the adoption of the standard Viterbi decoder. In fact, \ac{ML} decoding of a \ac{TB} \ac{CC} may be naively achieved by starting $M$ Viterbi decoders in parallel, where $M$ is the number of states in a trellis section. 
Each Viterbi decoder will have a different assumption on the starting state (that shall coincide with the final state due to the \ac{TB} constraint). The paths selected by the $M$ Viterbi decoders can be then used to form a list, within which lies the codeword that maximizes the likelihood $p\left(\by|\bx\right)$. This solution may become expensive from a computation viewpoint already for moderate-size encoder memories. A simple alternative to this approach is given by the \ac{WAVA} \cite{Fossorier03:WAVA}. The \ac{WAVA} is based on the recursive application of the Viterbi algorithm. In particular, one round of the Viterbi algorithm is applied to the \ac{TB} trellis at each iteration, using the final state probabilities computed in the past iteration as initial state probabilities, with the first round assuming the initial states to be equally likely. At the end of each iteration, the decoder checks if the selected path starts and ends in the same state (hence, fulfilling the \ac{TB} constraint). If the check is satisfied, then the decoder is stopped and the selected path is declared as final decision. Otherwise, another iteration of the Viterbi algorithm is performed. The process can be iterated for some preset maximum number of times. It turns out that, for many \ac{TB} \acp{CC}, four iterations are sufficient to attain near-\ac{ML} performance.

\begin{figure}[h]
	\begin{center}
		\includegraphics[width=0.9\textwidth]{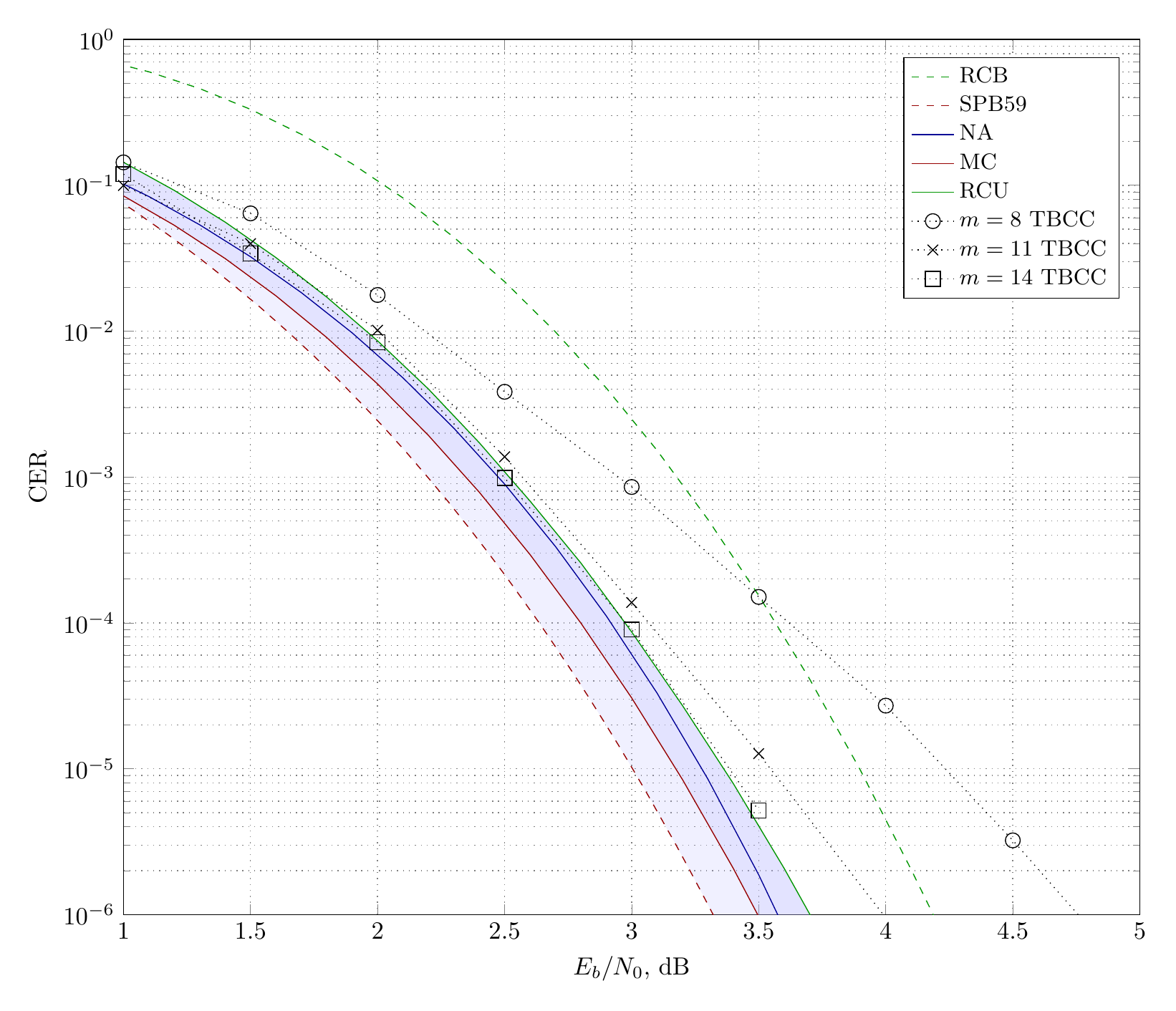}
	\end{center}
	\caption{Codeword error rate vs. $\snr$ for $(128,64)$ binary \ac{TB}\acp{CC} with different memory $m$, \ac{bi-AWGN} channel.}\label{fig:64128_TBCC}
\end{figure}

Figure~\ref{fig:64128_TBCC} shows the performance in terms of \ac{CER} vs. $\snr$ of $(128,64)$ binary \ac{TB}\acp{CC} with different memory $m$ and polynomials as specified in Table~\ref{tab:TBCCs}. For the case of $m=14$ the performance is within $0.07$ dB from the \ac{NA} at $\ac{CER}\approx 10^{-4}$. This results show that \ac{TB}\acp{CC} work very well for short blocks. 
Unfortunately, as we will see in Section~\ref{sec:comparison}, the memory must be increased as the blocklength grows in order to approach the finite-length bounds, rendering the scheme less practical.\footnote{For large memory, sequential decoding algorithms may be considered to reduce the decoding complexity. We refer the reader to \cite{FCC15}  for a thorough presentation of sequential decoders, including the advanced \ac{BEAST} algorithm of \cite{BEAST}.}

\subsection{CRC/TBCC Concatenation}

An alternative is the concatenation of CRC error-detection code with a punctured tailbiting convolutional code. Because the addition of the CRC code substantially increases coding overhead for short blocklengths, puncturing of the TBCC can be used to reduce the overhead back to the original level. Further, it is possible to jointly decode the CRC and TBCC codes so that the concatenated code error-correction performance and error-detection performance are both superior to that of the TBCC operating on its own.

One algorithm for decoding the \ac{CRC}/\ac{TB} \ac{CC} concatenation is the \ac{LVA} \cite{Sehadri94:LVA}. The \ac{LVA} would keep a list of the best paths to the termination node and choose the one that passed the CRC check. Of course, since the \ac{CC} is a \ac{TB} code, the algorithm would have to be a list version of the WAVA algorithm.

One may consider the concatenation of a CRC code with generator $g(x)$ and a \ac{CC} with generators $[g_1(x)\:\ g_2(x)]$ to be a catastrophic \ac{CC} (or encoder) with generators $[g(x)g_1(x)\:\ g(x)g_2(x)].$ In principle then, because the encoding will be terminated, one may decode the CRC/\ac{TB} \ac{CC} combination with a \ac{WAVA}. Of course, there will be a large number of states, making the decoder very complex. Even if the application allowed such a large complexity, this approach gives up the ability to reliably detect errors at the decoder output.

An algorithm that nicely trades off the error-correction and error-detection capabilities of the \ac{CRC}/\ac{TB} \ac{CC} will now be presented.\footnote{To our knowledge, this algorithm
has not been previously presented in the literature.} To simplify the presentation, we start with the assumption that there is a \ac{SISO} trellis decoder for the \ac{TB} \ac{CC} that has full knowledge of the starting and ending state of the \ac{TB} \ac{CC} encoder. We provide Algorithm \ref{alg:decode} with the necessary definitions given below:
\begin{itemize}
\item weak position $\triangleq$ unreliable position, bit position whose \ac{LLR} has small magnitude
\item $w$ $\triangleq$ current number of weak positions under test; hypothesized extrinsic information will be placed in these positions
\item $p$ $\triangleq$ current $w$-bit pattern being tested in $w$ weak positions
\item $\mcc{MaxWeak} \triangleq$ maximum value of $w$ (Typically, $3 \le \mcc{MaxWeak} \le 10$.)
\item int mask[$\mcc{MaxWeak}+1$] $\triangleq$ $\{1,2^1-1,2^2-1,2^3-1,\dots,2^{MaxWeak}-1 \}$
\item strong 1 $\triangleq$ large positive value (e.g., 100.0) used for extrinsic information
\item strong 0 $\triangleq$ large negative value (e.g., -100.0) used for extrinsic information
\item weakposn$[w][p]$ $\triangleq$ weakest position found after decoding with candidate (or hypothesized) $w$-bit pattern $p$ as strong 0's and 1's placed in the $w$ weak positions via extrinsic information
\item $p_0(b_0)$ $\triangleq$ value of bit $b_0$ in binary representation of integer $p_0$ (least significant bit is bit $0$)
\item strong $p_0(b_0)$ $\triangleq$ strong 1 if  $p_0(b_0)=1$, strong 0 if  $p_0(b_0)=0$
\end{itemize}

\begin{algorithm}
	\caption{High-level description of the decoding algorithm}
	\label{alg:decode}
	\begin{algorithmic}[1]
		\For{$w = 0,\dots,\mcc{MaxWeak}$}
		    \For{$p = 0,\dots,2^w-1$}
		        \If{$w>0$}
		            \State Set extrinsic information (see Algorithm \ref{alg:exInf})
		        \EndIf
		        \State \ac{SISO} decode (see Algorithm \ref{alg:SISODecode})
		        \If{\ac{CRC} fails (i.e., syndrome $\neq 0$)}
		            \State weakposn$[w][p]$ = weakest \ac{LLR} position
		        \Else
		            \State break
		        \EndIf
			\EndFor
		    \If{syndrome $== 0$}
		        \State break
		    \EndIf
		\EndFor
	\end{algorithmic}
\end{algorithm}

\begin{algorithm}
	\caption{\ac{SISO} Decode Step}
	\label{alg:SISODecode}
	\begin{algorithmic}[1]
		\State Apply the \ac{SISO} decoder, utilizing any extrinsic information provided, to the channel output word to provide \acp{LLR} for the \ac{TB} \ac{CC} encoder's input
		\If {the decoder output passes the \ac{CRC} check}
			\State Decoding is complete
		\Else
		    \State Note \ac{CRC} failure	
		\EndIf
	\end{algorithmic}
\end{algorithm}

\begin{algorithm}
	\caption{Set Extrinsic Information Step (for $w$-bit pattern $p$ in $w$ weak positions)}
	\label{alg:exInf}
	\begin{algorithmic}[1]
		\For{$b_0 = 0,\dots,w-1$}
				\State $p_0 = p$ \& mask[$b_0$]
				\State extrinsic[weakposn$[n_0][p_0]$] = strong $p_0(n_0)$
		\EndFor
	\end{algorithmic}
\end{algorithm}

Note that the decoder uses no extrinsic information the first time through
the outer for loop, after which the weakest LLR position is found (if CRC fails). The second time through the outer loop, strong 0 and then strong 1 extrinsic values are tested in the weakest position. The strong 1 is attempted only if the CRC fails when strong 0 is tested. New weakest positions are found after each strong value is attempted. The third time through the loop, two-bit patterns of strong 0's and strong 1's are attempted, each time checking the CRC and finding \mcc{the} newest weakest position if the CRC fails. The algorithm continues until there is a passed CRC event or the outer loop completes.

It should be clear from the algorithm that the larger the value of $\mcc{MaxWeak}$, the better the error-correction performance and the worse the error-detection performance. Note also, at low \ac{SNR} values, there can be up to $2^{\mcc{MaxWeak}+1}$ \ac{SISO} decodings---quite a large number. However, most applications with short blocklengths do not require decoding at high speeds. Also, as we shall see, the error-correction performance of this algorithm can be very good and it can be easily traded off with error-detection performance by decreasing $\mcc{MaxWeak}$.

The \ac{CRC} code we consider has generator polynomial $g(x)= x^{16}+x^{12}+x^5+1$. The \ac{TB} \ac{CC} we consider has generator polynomials $[g_1 \:\ g_2] = [5537, 6131]_{oct}$. With 64 input bits, the natural parameters for this \ac{CRC}/\ac{TB} \ac{CC} are  $(n,k)=(160,64)$. Consequently, to attain a $(128,64)$ code, we puncture every fifth bit of the encoder output, starting with the third bit.

The \ac{SISO} decoder employs the \ac{BCJR} algorithm. We consider two situations: (1) the \ac{TB} \ac{CC} encoder's starting and ending state is unknown to the \ac{SISO} decoder and (2) the starting and ending state is known to the \ac{SISO} decoder. For the first case, we use a \ac{WAVA}-like approach in the \ac{BCJR} decoder. We justify the second case by arguing that, in many applications, a packet number or an identification number is expected. Such a number can be moved to the end of the \ac{TB} \ac{CC} encoder input word so that the encoder starting and ending state is known.

\begin{figure}[h]
	\begin{center}
		\includegraphics[width=0.9\textwidth]{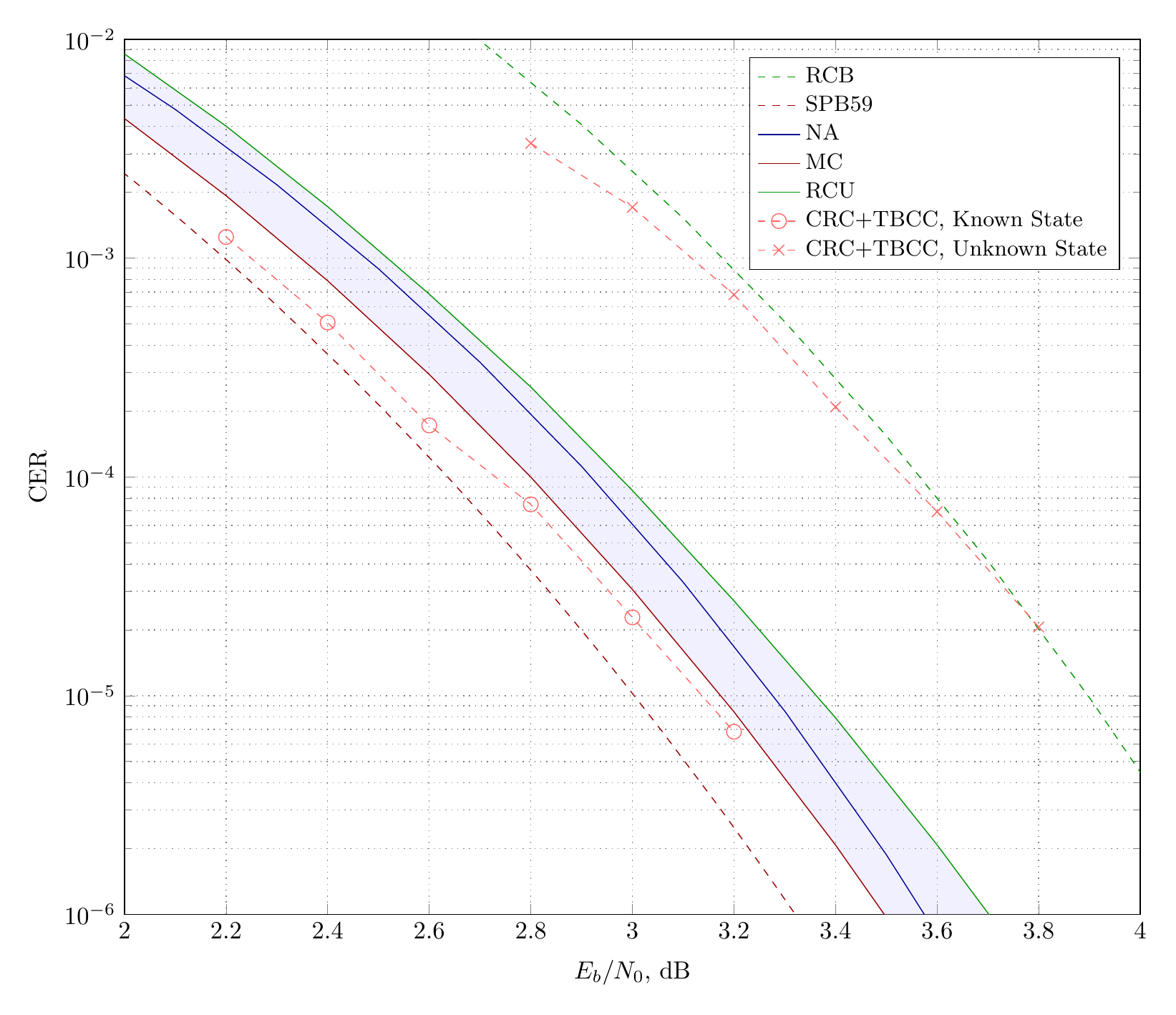}
	\end{center}
	\caption{Codeword error rate vs. $\snr$ for $(128,64)$ \ac{CRC}/\ac{TB} \ac{CC} concatenation over the \ac{bi-AWGN} channel.}\label{fig:64128_CRC_TBCC}
\end{figure}

Figure \ref{fig:64128_CRC_TBCC} plots the performance of the \ac{CRC}/\ac{TB} \ac{CC} under consideration on the \ac{bi-AWGN} channel for the unknown-state and known-state cases. Decoder parameter $\mcc{MaxWeak}$ was set to 10. As seen in the figure, the unknown-state case is superior to the turbo and \ac{LDPC} codes in Figure 5. Although it is unfair to compare the known-state case to the bounds, we see that the known-state \ac{CER} curve  is about 0.1 dB to the right of the SPB59 curve.

As for error-detection performance, we first define $P_{ud|e}$ to be the probability that an error at the decoder output is undetected by the decoder. For the simulation curves in the figure for which $\mcc{MaxWeak = 10}$, we measured $P_{ud|e}$ to be just under 0.1 for both cases. With $\mcc{MaxWeak = 4}$, we measured
$P_{ud|e}$ to be less than 0.001 for both cases. For $\mcc{MaxWeak = 4}$, the known-state \ac{CER} curve moves rightward about 0.4 dB and the unknown-state \ac{CER} curve moves rightward about 0.7 dB.

We point out that the \ac{CRC} and \ac{TB} \ac{CC} polynomials we chose were ``off the shelf" and it might be possible to design an improved \ac{CRC} code for a specific \ac{TB} \ac{CC} using the techniques in \cite{LouWesel15:CRC-CC}.

\FloatBarrier

\section{Modern Short Codes}\label{sec:codesmodern} 
In the following subsections, we briefly review some of the best constructions of modern channel codes for short blocklengths. The review includes both binary and non-binary turbo and \ac{LDPC} codes, as well as polar codes. 

\subsection{Binary Turbo and LDPC Codes}
\label{sec:codesmodern_bin_turbo_ldpc}

For short blocklengths, turbo and \ac{LDPC} codes are typically outperformed by the code classes discussed in the previous section. Their performance becomes competitive in the moderate blocklength regime thanks to their linear (in blocklength) decoding complexity for a fixed number of iterations. While this holds true for both binary and non-binary turbo/\ac{LDPC} codes, binary turbo and \ac{LDPC} codes retain a larger appeal from the perspective of decoder complexity.

\begin{figure}[h]
	\begin{center}
		\includegraphics[width=0.9\textwidth]{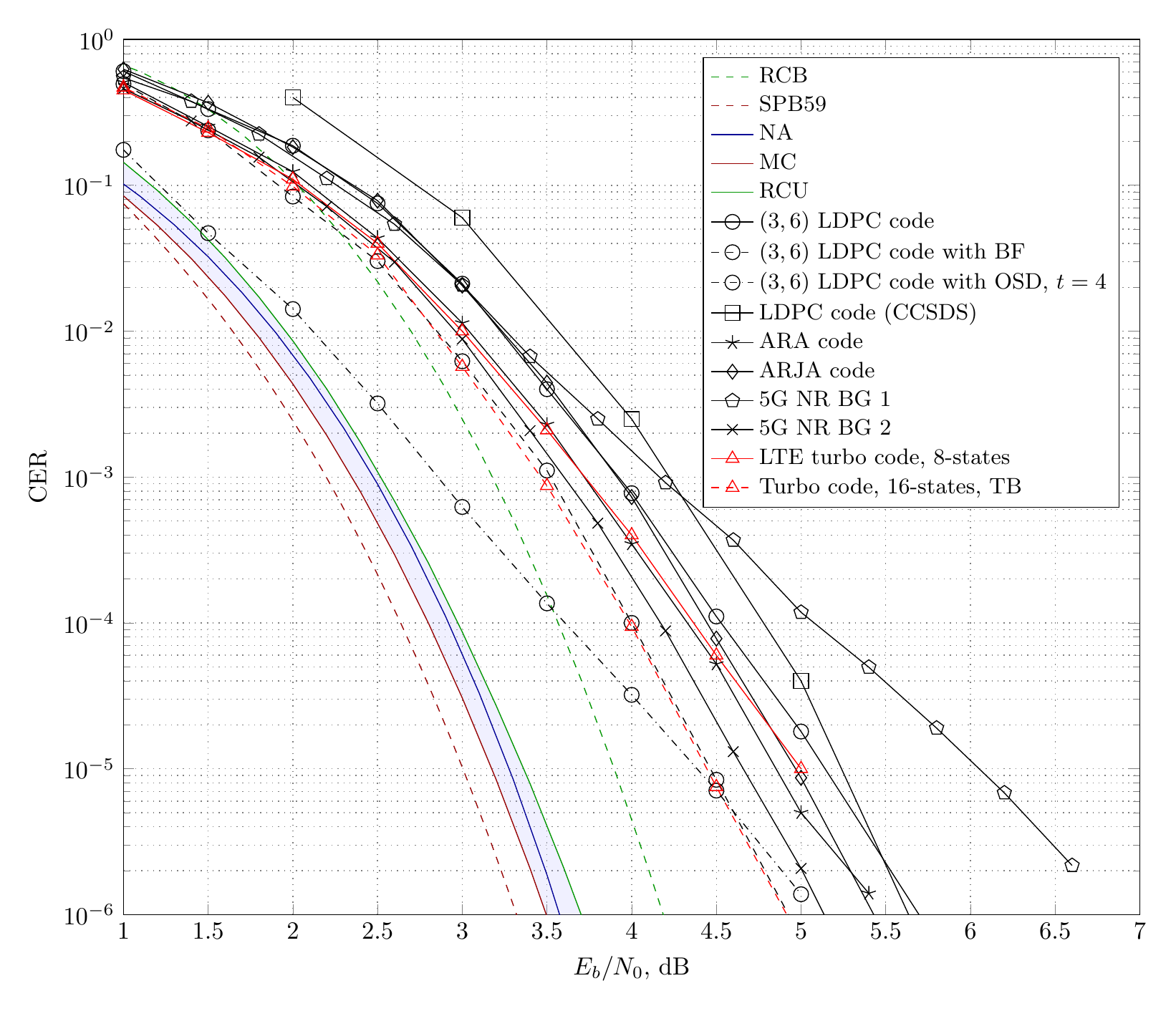}
	\end{center}
	\caption{Codeword error rate vs. $\snr$ for $(128,64)$ binary \ac{LDPC} and turbo codes, \ac{bi-AWGN} channel.}\label{fig:64128_TurboLDPC}
\end{figure}

Binary turbo codes \cite{Berrou93:TC} have been successfully included as a channel coding scheme in the 3G/4G cellular standards. Turbo codes are known to provide excellent coding gains in the moderate blocklength regime and (if carefully designed) at short blocklengths as well. If low error rates are required ($\CER<10^{-4}$), a convenient design choice is to adopt $16$-state component codes, i.e., to use memory-$4$ convolutional codes in the parallel concatenation, together with \ac{TB} termination for the component codes \cite{Crozier01:DRP}. The small size of the information word permits an efficient interleaver optimization. Code-matched  \cite{WJV02} and protograph-based \cite{garzon2018protograph} interleavers in particular turn out to be very effective in lowering error floors. The performance of two $(128,64)$ turbo codes with memory-$3$ and memory-$4$ component codes is provided in Figure \ref{fig:64128_TurboLDPC}. The first code is from the \ac{LTE} standard, whereas the second code has been designed with the interleaver construction of \cite{Jerkovits16:Short} and exploits \ac{TB} component codes. The second code performs fairly close to the \ac{RCB}, and nearly $1$ dB away from the \ac{NA} at $\CER=10^{-4}$. The \ac{LTE} turbo code loses almost $0.4$ dB at the same target \ac{CER}. Remarkably, the simple $16$-state construction provides a performance that is among the best achievable by binary iteratively-decodable codes, at least down to moderate error rates.

\ac{LDPC} codes \cite{Gallager63:LDPC} are particularly attractive thanks to their excellent performance and to the possibility of developing high-throughput iterative decoders based on the codes' Tanner graph \cite{tanner} with a large degree of parallelism. 
\ac{LDPC} codes can be subdivided into two broad categories: unstructured and structured \ac{LDPC} codes \cite{Richardson08:BOOK,RyanLin09:BOOK}. For an unstructured \ac{LDPC} code, the code parity-check matrix is designed according to a  computer-based pseudo-random algorithm that places the non-zero entries (aiming, for instance, at maximizing the girth of the corresponding Tanner graph \cite{Hu05:PEG}). Unstructured \ac{LDPC} codes are rarely implemented in practice~\cite{Thorpe03:PROTO}.
Among structured \ac{LDPC} codes, protograph-based codes \cite{Thorpe03:PROTO,Multi_EDGE} are particularly interesting from a decoder implementation viewpoint. A protograph  is a relatively small graph from which a larger Tanner graph
can be obtained by a copy-and-permute procedure: the protograph is copied $Q$
times, and then the edges of the individual replicas are permuted among the
replicas (under some restrictions described in \cite{Thorpe03:PROTO}) to obtain a single, large graph. The parameter $Q$ is often referred to as a \emph{lifting factor}.

When cyclic edge permutations are used, the code associated with the Tanner graph is quasi-cyclic, facilitating the implementation of efficient encoders and decoders \cite{RyanLin09:BOOK,Mansour}. Powerful protograph \ac{LDPC} codes have been designed during the past decade \cite{Divsalar09:Proto}. 
A class of protograph \ac{LDPC} codes that performs remarkably well down to short blocklengths is that of the \ac{ARA} codes \cite{Divsalar07:ARA}. The performance of an $(128,64)$ \ac{ARA} code is provided in Figure \ref{fig:64128_TurboLDPC}. The code performs close to the \ac{LTE} turbo code. An error floor appears at a \ac{CER} below $10^{-5}$. The performance of an $(128,64)$ \ac{LDPC} code based on a slightly modified protograph, dubbed \ac{ARJA} \cite{Divsalar09:Proto}, is provided too. The \ac{ARJA} code trades a negligible loss in the waterfall region for a superior performance at large \acp{SNR}, i.e., it has a lower error floor.

Another class of protograph \ac{LDPC} codes with excellent performance is the one proposed in \cite{Chen15:ProtoRaptor}, which relies on the concatenation of an outer high-rate \ac{LDPC} code with an inner  \ac{LDPC} code. The inner \ac{LDPC} code construction  resembles an LT code \cite{luby02:LT}, resulting in an overall \ac{LDPC} code structure that closely mimics that of a Raptor code \cite{Shokrollahi06:raptor} (the main difference is that here the bits at the input of the inner LT encoder are, with the exception of the punctured ones, sent over the channel). This design paradigm has been adopted in the 5G standard \cite{Richardson5G}. 

In particular, the upcoming 5G \ac{NR} standard foresees the use of two protograph-based codes for its \ac{eMBB} use case. Their design reflects the requirements for 5G NR, which includes the support of a wide range of blocklengths and code rates and a {naive} integration of \ac{H-ARQ}. Additionally, the nested structure of the codes and the quasi-cyclic lifting allow a hardware-friendly implementation with minimal description complexity as well as various possibilities for parallelization. 
Base graph\footnote{In the 5G \ac{NR} jargon, base graph is synonymous with protograph.} 1 (BG 1) targets larger blocklengths and higher rates ($500 \leq k \leq 8448$, $1/3 \leq R \leq 8/9$), whereas base graph 2 (BG 2) is optimized for smaller blocklengths and lower rates ($40 \leq k \leq 2560$, $1/5 \leq R \leq 2/3$). Both base graphs make use of punctured variable nodes. This construction is known to significantly improve the decoding threshold~\cite{Divsalar09:Proto}.
We observe in Figure \ref{fig:64128_TurboLDPC} that the $(128, 64)$ 5G \ac{NR} \ac{LDPC} code based on BG 2 even slightly outperforms the \ac{ARA} code with the same code parameters. In contrast, the performance of an \ac{LDPC} code constructed from BG 1 (which is optimized for larger blocklengths and higher code rates) is severely degraded due to its poor minimum distance.

As described in Section \ref{sec:codesclassic:OSD}, a conceptually simple improvement to the \ac{BP} decoder performance can be obtained by applying \ac{OSD} whenever the \ac{BP} decoding fails to converge to a valid codeword. In Figure \ref{fig:64128_TurboLDPC}, we provide the performance of $(3,6)$ regular \ac{LDPC} code under the \ac{BP} decoder followed by an additional \ac{OSD} (with order $t=4$) step applied whenever the \ac{BP} decoder fails after a maximum number of $50$ iterations. The performance gain over iterative decoding is around $1$ dB at \cer{-4}. However, the gap reduces to $0.5$ dB at \cer{-5}.
Besides \ac{OSD}, another list decoding algorithm that can improve remarkably the performance of short (binary) \ac{LDPC} codes is the bit flipping (BF) algorithm proposed in \cite[Algorithm 8.6]{Ostojic2010}.
In Figure \ref{fig:64128_TurboLDPC}, we depict the performance of this algorithm when applied to the $(3,6)$ regular \ac{LDPC} code, for the case of a maximum number of $200$ bit flips. The gain over iterative decoding exceeds $0.5$ dB at \cer{-5}.

\subsection{Non-Binary Turbo and LDPC Codes}\label{sec:codesmodern:nb} 

Turbo codes constructed over non-binary finite fields were originally investigated in \cite{Berkmann00:Diss}. In \cite{Liva13:ShortTC}, a design based on memory-$1$ (in terms field elements) time-varying recursive \ac{TB} encoders was proposed, which yields among the best known performance for iteratively-decodable short codes down to very low error rates. The construction is  particularly effective for relatively large finite fields (e.g., $\ff{64}$ and $\ff{128}$). The component-code \ac{BCJR} decoder can be efficiently implemented by means of the \ac{FFT} \cite{Berkmann02:dualtrellis,Liva13:ShortTC}, yielding remarkable savings in complexity (although the decoding complexity remains considerably larger than that of a binary turbo code). Further efficient decoder implementations have been recently investigated in \cite{klaimi2018low} showing how most of the coding gains can be preserved even when dramatically reducing the decoding complexity.

Non-binary \ac{LDPC} codes \cite{Davey98:NonBinary} based on ultra-sparse parity-check matrices \cite{Poulliat08:BinImag} match tightly the performance of non-binary turbo codes, down to very low error rates, when constructed on finite fields of order larger than or equal to $64$ \cite{Liva12:CCSDS,Divsalar12:NonBinaryShort}. 
In fact, it was shown in \cite{Liva13:ShortTC} that non-binary turbo codes based on memory-$1$ time-varying recursive \ac{TB} encoders admit a simple protograph \ac{LDPC} representation, and correspond to a special class of non-binary ultra-sparse \ac{LDPC} codes.  While the decoding of non-binary \ac{LDPC} codes can be largely simplified by employing the \ac{FFT} at the check nodes (with probability-domain decoding), efficient implementations in the log-domain are still an area of active research \cite{Declercq07:FFT}. 

\begin{figure}[h]
	\begin{center}
		\includegraphics[width=0.9\textwidth]{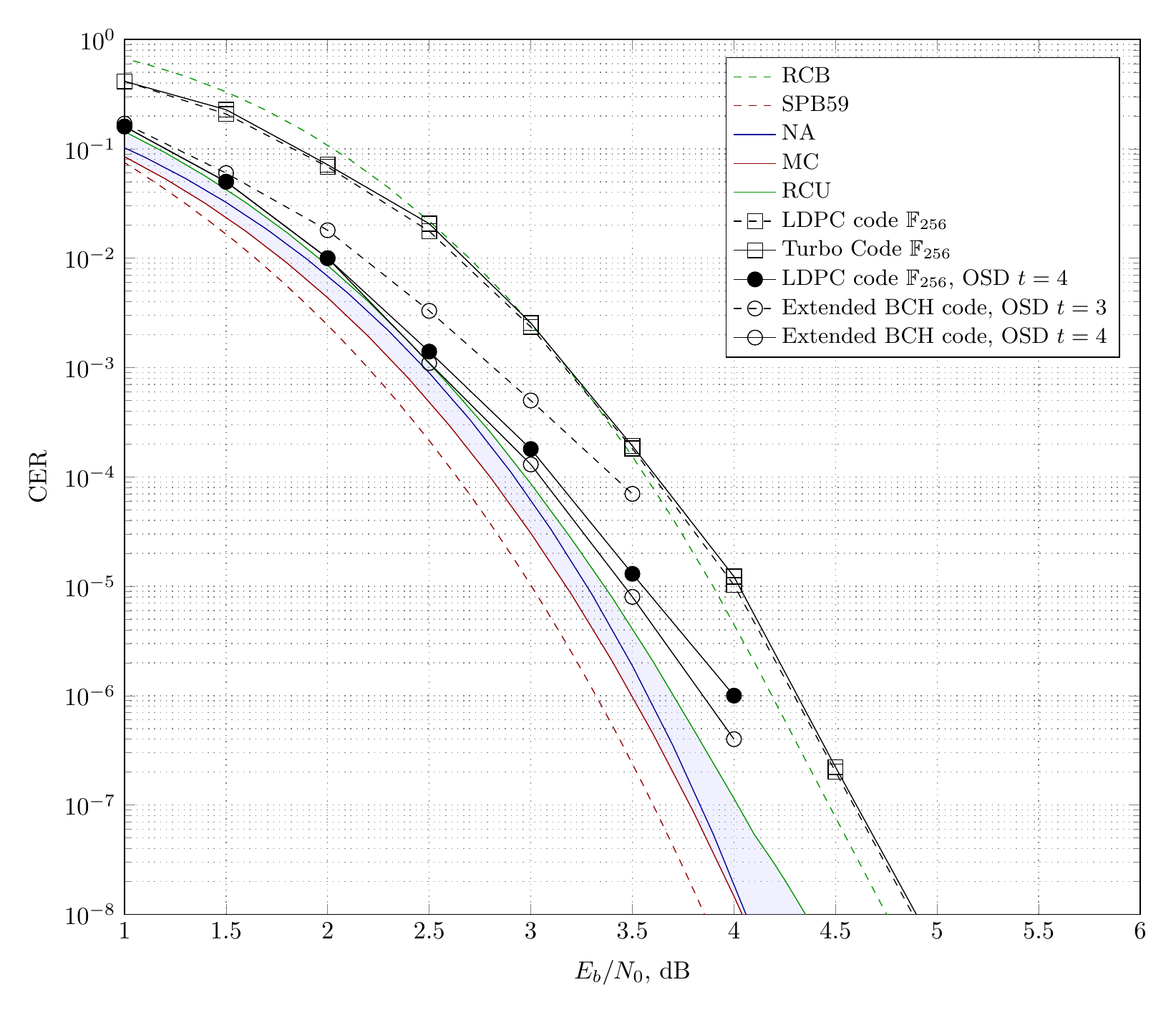}
	\end{center}
	\caption{Codeword error rate vs. $\snr$ for $(128,64)$ \ac{LDPC} and turbo codes over $\ff{256}$, \ac{bi-AWGN} channel.}\label{fig:64128_NB}
\end{figure}

Figure \ref{fig:64128_NB} shows the performance of non-binary turbo/\ac{LDPC} codes with block length $128$ and dimension $64$ on the \ac{bi-AWGN} channel. Both codes are constructed on $\ff{256}$. The \ac{LDPC} code has been considered for standardization within \ac{CCSDS} (as error-correcting code for satellite telecommand) \cite{Liva12:CCSDS,CCSDS14,CCSDS15} and it has been designed according to the method proposed in \cite{Poulliat08:BinImag}. The turbo code has been designed according to the method proposed in \cite{Liva13:ShortTC}. Both codes perform almost identically down to very low error rates, almost matching the \ac{RCB}.

Also for the non-binary case, a further decoding step based on \ac{OSD} decoding can be applied to any iteratively-decodable code whenever the \ac{BP} decoder fails. As an example, Figure \ref{fig:64128_NB} reports the performance of a $(128,64)$ non-binary \ac{LDPC} code constructed on $\ff{256}$ on the \ac{bi-AWGN} channel. After iterative decoding, when the decoder output does not fulfill the code parity-check equations, an additional \ac{OSD} step is applied with $t$ set to $4$. 
Specifically, \ac{OSD} is applied to the binary image of the non-binary \ac{LDPC} code. The performance is very close to the one attained by the $(128,64)$ extended \ac{BCH} code, first presented in Figure \ref{fig:64128_BCH}, gaining $0.5$ dB over the code performance under \ac{BP} decoding. Considering as reference an \ac{SNR} of $3$ dB, the \ac{BP} decoder for the non-binary \ac{LDPC} code fails with a probability close to $2\times 10^{-3}$. Hence, the \ac{OSD} is effectively activated only for a very small fraction of the transmissions. 

\subsection{Polar Codes}
\label{sec:polar}

Polar codes \cite{stolte2002rekursive,Arikan09:polarization} are the first class of provably capacity-achieving codes with low encoding/decoding complexity over any symmetric \ac{B-MC} under \ac{SC} decoding \cite{Arikan09:polarization}. The underlying idea behind polar codes, called channel polarization, is to take the independent copies of a symmetric \ac{B-MC} and convert them into noiseless and useless synthetic channels by applying a transform to input bits and by imposing a decoding order so that coding becomes trivial: Transmit information bits over the noiseless synthetic channels while inputs to the useless ones are set (frozen) to a predetermined value, e.g., to $0$, and the decoder knows these bits before transmission. Those input bits are called frozen bits. 
As the number of polarization steps grows, the fraction of noiseless synthetic channels tends to the channel capacity, while the fraction of useless channels tends to its complement to one. 

\begin{figure}[h]
	\begin{center}
		\includegraphics[width=0.9\textwidth]{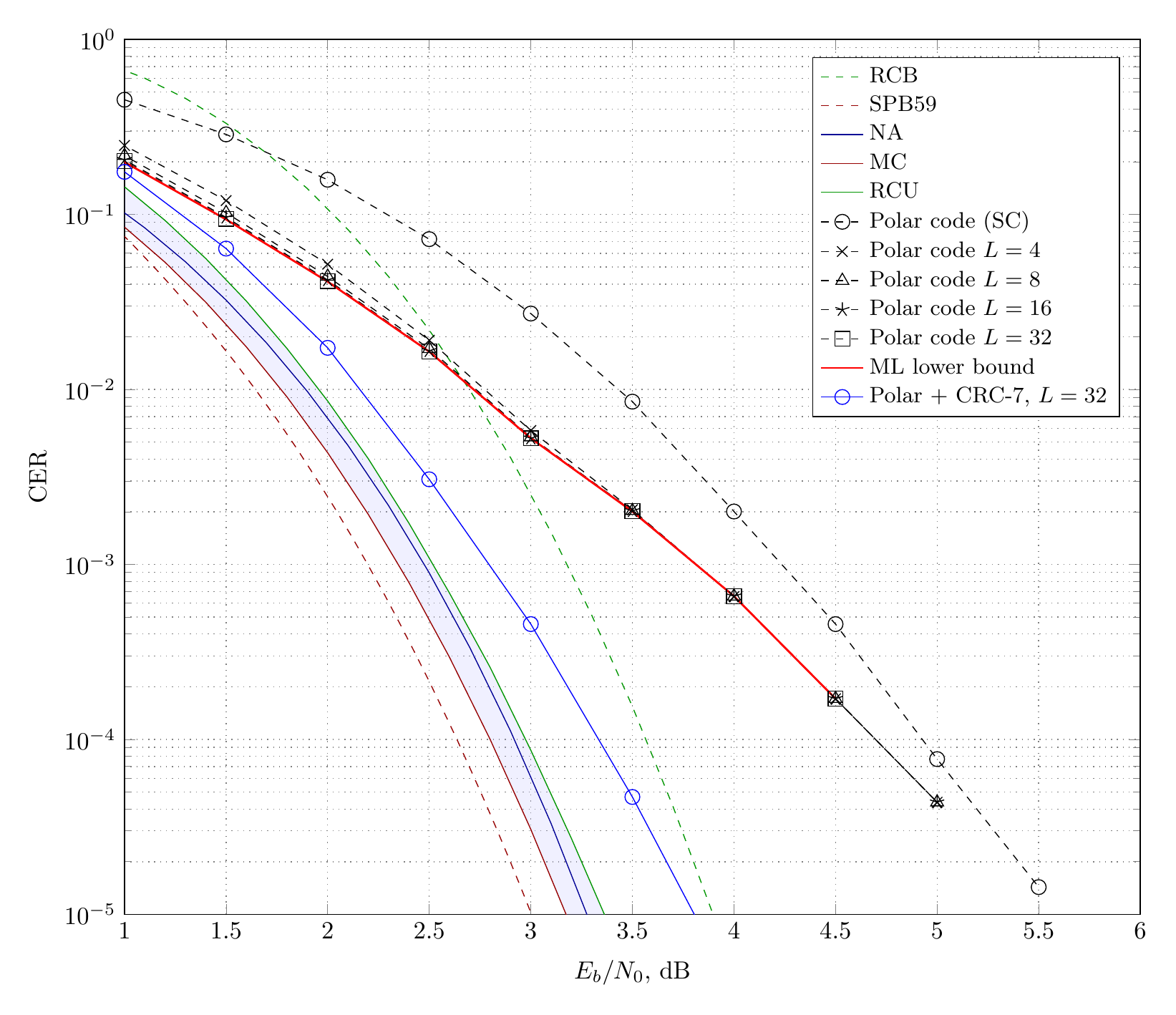}
	\end{center}
	\caption{Codeword error rate vs. $\snr$ for a $(128,64)$ polar code, \ac{bi-AWGN} channel.}\label{fig:64128_polar}
\end{figure}

For a given \ac{SNR}, constructing an $(n,k)$ polar code requires one to find the least reliable $n-k$ synthetic channels, or equivalently, bit positions. The design is not universal, i.e., the polar code design differs depending on the channel quality. Monte Carlo-based designs were proposed in \cite{Arikan09:polarization,stolte2002rekursive}, while a \mcc{density evolution}-based construction is introduced in \cite{Mori09}. An efficient implementation for density evolution is provided in \cite{Tal:HowToConstruct} together with an analysis providing lower and upper bounds for the reliabilities of the bit positions. The \ac{GA} for design was proposed in~\cite{trifonov_efficient_2012}. Other methods based on a partial order among the positions  were proposed in \cite{Mondelli17,he_beta-expansion:_2017}. These methods allow one to design frozen bit sequences that show a good behavior for a wide range of channel parameters and rates. This has been of particular importance during 5G standardization~\cite{bioglio_design_2018} with its strong emphasis on lowering the description complexity.

Although being capacity-achieving under \ac{SC} decoding, the effectiveness of polar codes for short blocklengths comes only after modifying both the decoder and the code, i.e., by employing the \ac{SCL} decoder of \cite{Tal15:ListPolar} aided by the addition of an outer high-rate code (typically, a \ac{CRC} code). In fact, in the short and moderate-blocklength regimes the performance of \ac{SC} decoding of polar codes falls short of the performance under \ac{ML} decoding. In the \ac{SCL} decoding algorithm, a set of \ac{SC} decoders work in parallel producing a list of different codeword candidates for the same observed channel output. The complexity of the algorithm is linear in the list size. \mcc{The outer high-rate code, which improves the distance properties of the resulting code, is used to test the list of codewords produced by the \ac{SCL} decoder.} Among the survivors, the one with the largest likelihood is picked as the decoder output. \mcc{The design of the concatenated code for \ac{SCL} decoding becomes more sophisticated due to the increased search space, i.e., unmanageable number of possible interleavers and outer codes \cite{TSE:RMPolar14,RBCL17,RBC18,Yuan:ListConst19}}.

The performance of a $(128,64)$ polar code designed by using the \ac{GA} of \ac{DE} for the \ac{bi-AWGN} channel with \mcc{$E_b/N_0=4.5$} dB under \ac{SC} and \ac{SCL} decoding is shown in Figure \ref{fig:64128_polar}. By increasing the list size, close-to-\ac{ML} performance is achieved. In fact, a lower bound on the \ac{ML} error probability can be obtained by artificially introducing the correct codeword in the final list, prior to the final selection. 
One can see from the figure that the lower bound on the \ac{ML} error probability is approached quickly as the list size $L$ grows. Already for $L=8$, the gap from the the \ac{ML} lower bound is nearly invisible for the setup considered in the figure.
The performance of the concatenation of a $(128,71)$ polar code with a \ac{CRC}-$7$ code as an outer code is shown as well. The inner polar code was designed for  \mcc{$E_b/N_0=5$} dB. The outer \ac{CRC} code has generator polynomial $g(x)=x^7 + x^3 + 1$, leading to a code with dimension $64$. 
A list size of $32$ has been used in the simulation. The code performs remarkably close to the \ac{RCU} bound down to low error rates.

\section{Code Comparison: Examples}\label{sec:comparison}

\subsection{Very Short Codes}

\begin{figure}[h]
	\begin{center}
		\includegraphics[width=0.9\textwidth]{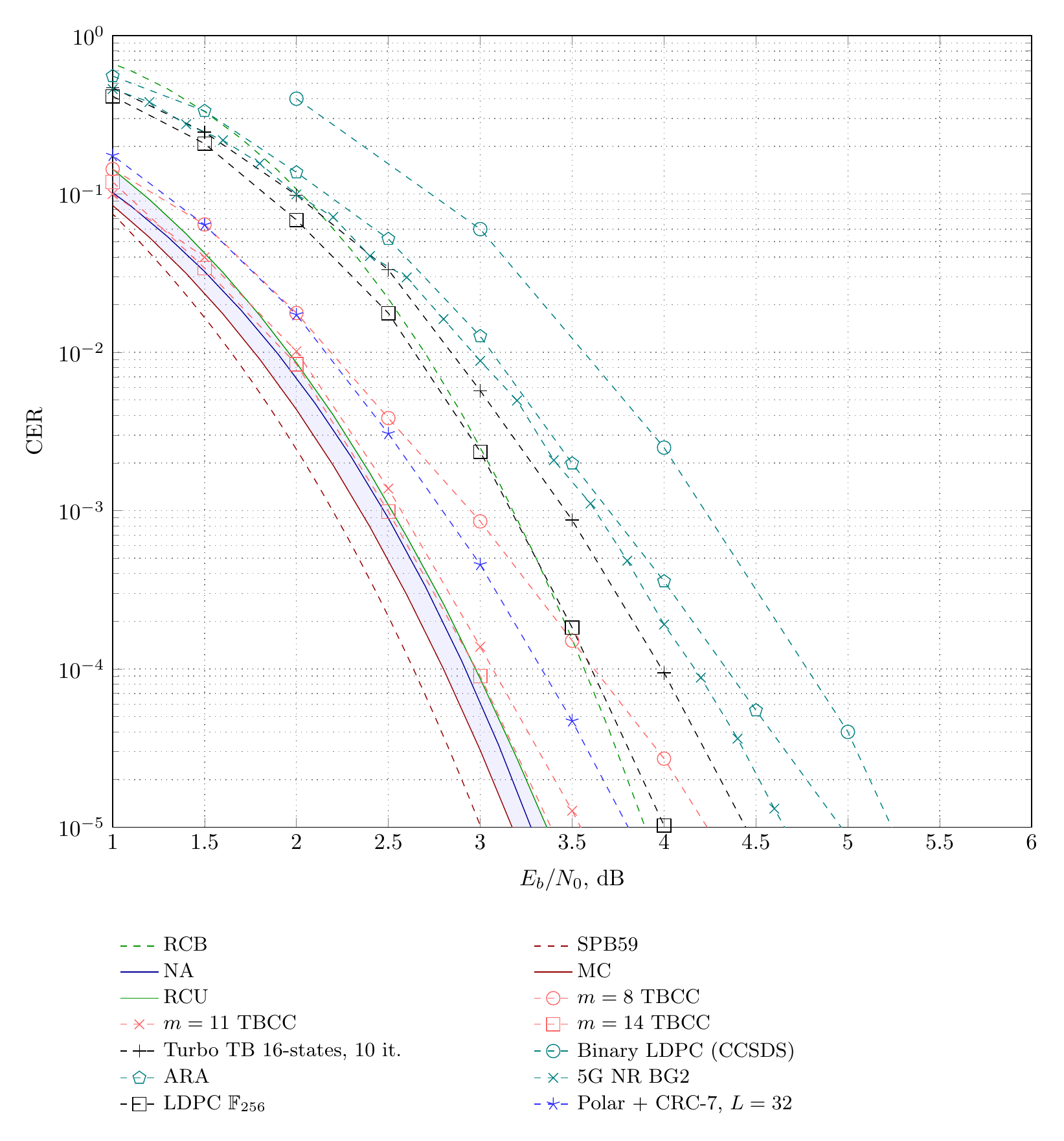}
	\end{center}
	\caption{Codeword error rate vs. $\snr$ for several $(128,64)$ codes over the \ac{bi-AWGN} channel.}\label{fig:64128}
\end{figure}

In this section, we summarize the results reported in the previous sections about the performance of very short codes over the \ac{bi-AWGN} channel. We focus on  codes with blocklength $n=128$ and code dimension $k=64$ bits.
The performance of the codes is compared in Figure \ref{fig:64128}. As reference, the performance of the $(128,64)$ binary protograph-based \cite{Divsalar07:ShortProto} \ac{LDPC} code from the \ac{CCSDS} telecommand standard \cite{CCSDS14} is provided too. The \ac{CCSDS} \ac{LDPC} code performs somehow poorly in terms of coding gain and is outperformed by the \ac{ARA} \ac{LDPC} code.\footnote{All \ac{LDPC} codes considered in this section have been designed by means of a girth optimization  based on the \ac{PEG} algorithm \cite{Hu05:PEG}. A maximum of $200$ belief propagation iterations have been used in the simulations (although the average iteration count is much lower, especially at high \acp{SNR}, thanks to early decoding stopping rules).} At low error rates (e.g. \cer{-6}) the \ac{CCSDS} \ac{LDPC} code is likely to attain lower error rates than the \ac{ARA} code thanks to its remarkable distance properties \cite{Divsalar07:ShortProto}. 
Among the \ac{LDPC} codes adopted for the 5G \ac{NR} standard, the codes based on BG 2 are seen to be competitive, outperforming the \ac{ARA} code.

The performance of a  turbo code introduced in \cite{Garello13} based on $16$-state component recursive convolutional codes is also provided. The turbo code shows superior performance with respect to binary \ac{LDPC} codes, down to low error rates. The code attains a \cer{-4} at almost $0.4$ dB from the \ac{RCB}. The code performance diverges remarkably from the \ac{RCB} at lower error rates, due to the relatively low code minimum distance. Results for a non-binary \ac{LDPC} code are included in Figure \ref{fig:64128}. The code has been constructed over $\ff{256}$, and it attains visible gains with respect to its binary counterparts, performing on top of the \ac{RCB} (and $0.7$ dB away from the \ac{NA}) down to low error rates (no floors down to \cer{-9} were observed in \cite{Liva12:CCSDS}).
The error probability of the polar-code concatenation using a CRC-$7$ as an outer code is shown. The polar code has parameters $(128,71)$. A list size of $32$ has been used in the simulation. The code outperforms all the competitors that rely on iterative decoding algorithms.
Finally, the \ac{CER} of three \ac{TB} \acp{CC} is included \cite{Stahl99:OptimumTB,Gaudio2017}. The three codes have memory $8$, $11$ and $14$, respectively. Their generators (in octal notation) and their distance properties are summarized in Table \ref{tab:TBCCs}. The \ac{WAVA} algorithm has been used for decoding \cite{Fossorier03:WAVA}. The memory-$11$ convolutional code reaches the performance of the \ac{BCH} and \ac{LDPC} codes under \ac{OSD}.
The memory-$8$ code loses $1$ dB at \cer{-5}, but still outperforms binary \ac{LDPC} and turbo codes over the whole simulation range. The third code  (memory-$14$)  outperforms all other codes in Figure \ref{fig:64128} (at the expense of a high decoding complexity due to the large number of states in the code trellis).

\begin{table}[h]
	\centering
	\caption{Summary of the \ac{TB} \acp{CC} used in the comparisons.}
	\label{tab:TBCCs}
	\begin{tabular}{cccl}  
		\toprule
		\toprule
		Generators & $m$ & $ (n,k) $ & Weight enumerating function $A(\dumx) $ \\
		\midrule
		$ [515, 677] $ & $8$ & \multirow{3}{*}{$ (128,64) $} & $ 1 + 576\dumx^{12} + 1152\dumx^{13} + 1856\dumx^{14} +\ldots$ \\
		$ [5537, 6131] $ & $11$ & & $ 1 + 64\dumx^{14} + 960\dumx^{15} + 1356\dumx^{16} +\ldots$ \\ 
		$ [75063, 56711] $ & $14$ & & $ 1 + 8\dumx^{16} + 1856\dumx^{18} + 19392\dumx^{20}  +\ldots$ \\ \cmidrule{1-4}
		$ [515, 677] $ & $8$ & \multirow{3}{*}{$ (512, 256) $} & $ 1 + 2304\dumx^{12} + 4608\dumx^{13} + 7424\dumx^{14} + \ldots $ \\
		$ [5537, 6131] $ & $11$ & & $ 1 + 256\dumx^{14} + 3840\dumx^{15} + 5376^{16} + \ldots$ \\
		$ [75063, 56711] $ & $14$ & & $ 1 + 6656\dumx^{18} + 42240\dumx^{20} + 216320\dumx^{22} + \ldots$ \\
		\bottomrule 				
		\bottomrule
	\end{tabular}
\end{table}

\FloatBarrier

\subsection{Moderate-length Codes}

\begin{figure}[h]
	\begin{center}
		\includegraphics[width=0.9\textwidth]{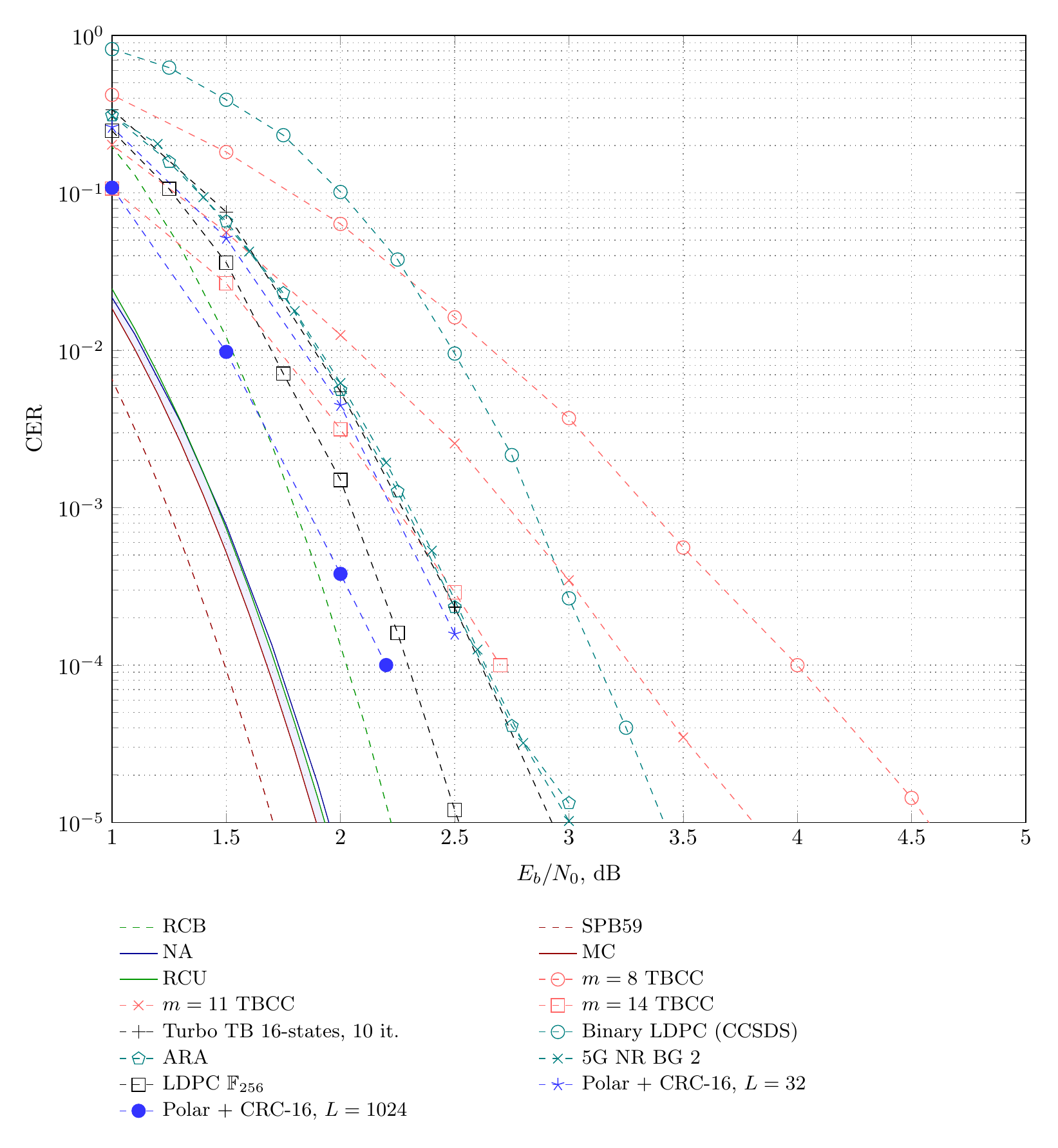}
	\end{center}
	\caption{Codeword error rate vs. $\snr$ for several $(512,256)$ codes over the \ac{bi-AWGN} channel.}\label{fig:256512}
\end{figure}

\begin{figure}[h]
	\begin{center}
		\includegraphics[width=0.9\textwidth]{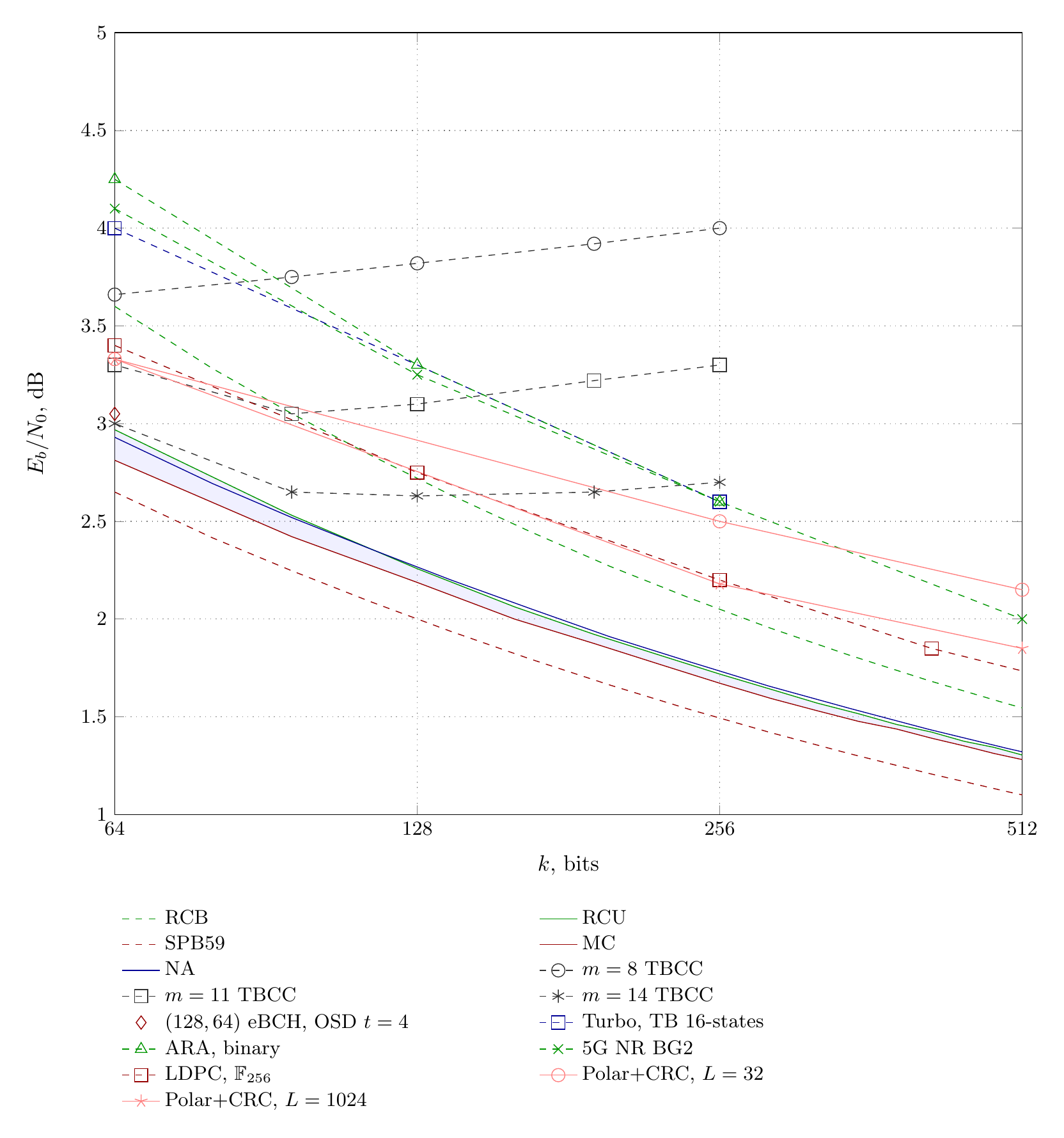}
	\end{center}
	\caption{$E_b/N_0$ required to achieve a codeword error rate of $10^{-4}$ for several code families over the \ac{bi-AWGN} channel. Here $R=1/2$.}\label{fig:CC}
\end{figure}

In this section, we address a second case study, where an intermediate blocklength of $n = 512$ bits is considered. The code dimension is fixed to $k=256$ bits yielding a rate $R=1/2$.
The performance of the codes is compared in Figure \ref{fig:256512} for transmission over the \ac{bi-AWGN} channel. 
Also here, the performance of the $(512,256)$ binary protograph-based \cite{Divsalar07:ShortProto} \ac{LDPC} code from the \ac{CCSDS} telecommand standard \cite{CCSDS14} is provided as a reference.
Most of the considerations that are valid in the very short blocklength regime are still valid here, with a few notable exceptions. First, we  observe that the performance of the polar code (concatenated with an outer $16$ bits \ac{CRC} code) is still competitive, but it performs only marginally better than binary \ac{LDPC} and turbo codes when the list size is limited to $32$. To close the gap to the finite length bounds, a larger list size (e.g., $1024$) has to be used. A second major discrepancy with respect to the very short block regime deals with the performance of \ac{TB} \acp{CC}. For the code parameters considered in Figure \ref{fig:256512}, \ac{TB} \acp{CC} are far from the finite length bounds even for the memory-$14$ case. This is an instance of a well known limitation of (\ac{TB}) \acp{CC}, i.e., the saturation, for large enough $n$, of the  \ac{TB} \ac{CC} minimum distance to the free distance of the underlying (unterminated) convolutional code (in addition, the minimum weight multiplicity grows with $n$). This phenomenon is illustrated in Figure~\ref{fig:CC}, where the \ac{SNR} required to achieve a target $\CER=10^{-4}$ is provided as a function of the code dimension $k$, for various code families.

\FloatBarrier

\subsection{Short Codes in Coded Modulation Schemes}\label{sec:coded_modulation}

Higher-order modulation increases the \ac{SE} of a communication system by using constellations with more than two signal points (e.g., $M$-amplitude shift keying (ASK) or $M$-quadrature amplitude modulation (QAM)) and transmitting more than one bit per channel use~\cite{ungerboeck_channel_1982}. As this requires an
interplay of both modulation and coding techniques, the term ``coded modulation'' (CM) has been established.
The most straightforward CM approach combines an $M$-ary constellation with a non-binary channel code over a field of order $M$.\footnote{It is also possible to map sequences of constellation points to one Galois field symbol of appropriate field order.} In this case, symbol-metric decoding (SMD) can be employed at the receiver. This is the common approach for non-binary LDPC and Turbo codes.

Practical receivers resort to ``pragmatic'' CM schemes with binary channel codes.
In such pragmatic schemes,  an $m$-bit binary labeling is assigned to each of the $M=2^m$ constellation points (e.g., a \ac{BRGC} code~\cite{gray1953pulse}) and a bit-wise decoding (BMD) metric is used at the decoder. The BMD metric is obtained by marginalizing over all bit levels except the one of interest, which causes a performance loss compared to SMD. This loss is particularly pronounced at low code rates. The best known example of pragmatic CM scheme is bit-interleaved coded modulation (BICM)~\cite{zehavi_8-psk_1992,guillen_i_fabregas_bit-interleaved_2008}. Binary LDPC and turbo codes are commonly combined with higher-order modulations using BICM. Polar codes achieve superior performance with multi-level coding and multistage decoding~\cite{imai_MLC,seidl_polar-coded_2013,Boch1703:Efficient} due to the improved polarization process.

The use of ASK/QAM constellations with uniformly distributed constellation points incurs a performance degradation for the AWGN channel, which is known as \textit{shaping loss}. Recently, many research efforts have focused on geometric and probabilistic shaping (GS/PS) approaches to overcome this deficit~\cite{sun_approaching_1993,bocherer_bandwidth_2015} and close the gap to the Shannon limit. Simulation results \cite{Stei1702:Comparison} show that PS signaling generally performs better than GS for the same constellation size. 
Additionally, PS allows a fine granularity in \ac{SE} as it can be tuned by means of a distribution matcher (DM)~\cite{schulte_constant_2016} and does not require different modulation and code rate combinations. 
To implement PS with coding, \ac{PAS} was proposed~\cite{bocherer_bandwidth_2015}, which circumvents the drawbacks of previous approaches (e.g., error propagation and the need for iterative demapping as a result of a one-to-many mapping). \ac{PAS} uses a shaping encoder before the  encoder (reverse concatenation), and a systematic generator matrix for encoding to maintain the desired distribution.
Furthermore, it exploits the symmetry of the capacity achieving distribution of the Gaussian channel.

\begin{figure}[h]
\centering
\includegraphics[width=0.9\textwidth]{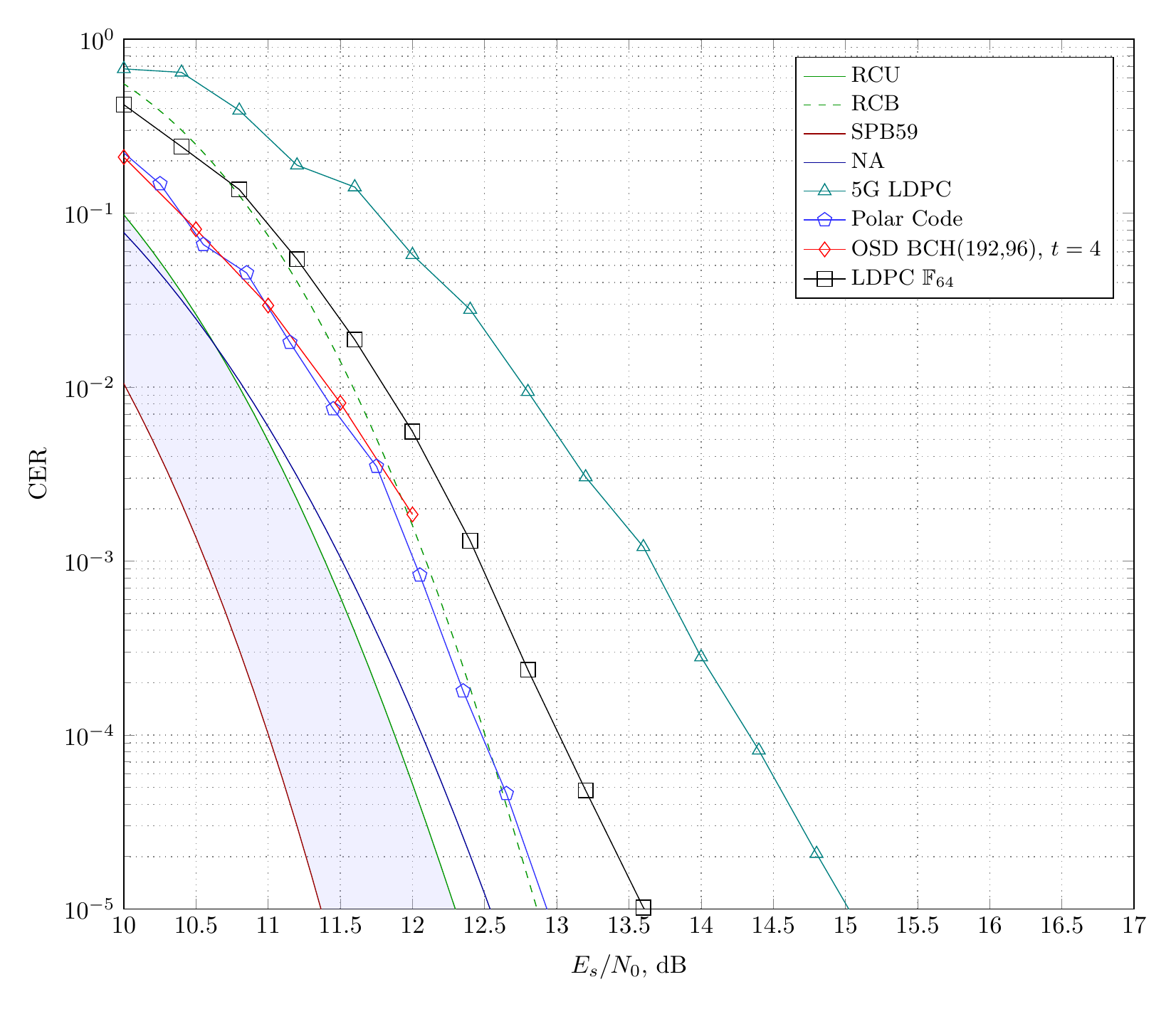}
\caption{Codeword error rate vs. $E_s/N_0$ for $64$-QAM, uniform signaling and an \ac{SE} of \SI{3.0}{bpcu} for several codes over the AWGN channel. The number of channel uses is $32$.}
\label{fig:cmp_192_uni}
\end{figure}

\begin{figure}[h]
\centering
\includegraphics[width=0.9\textwidth]{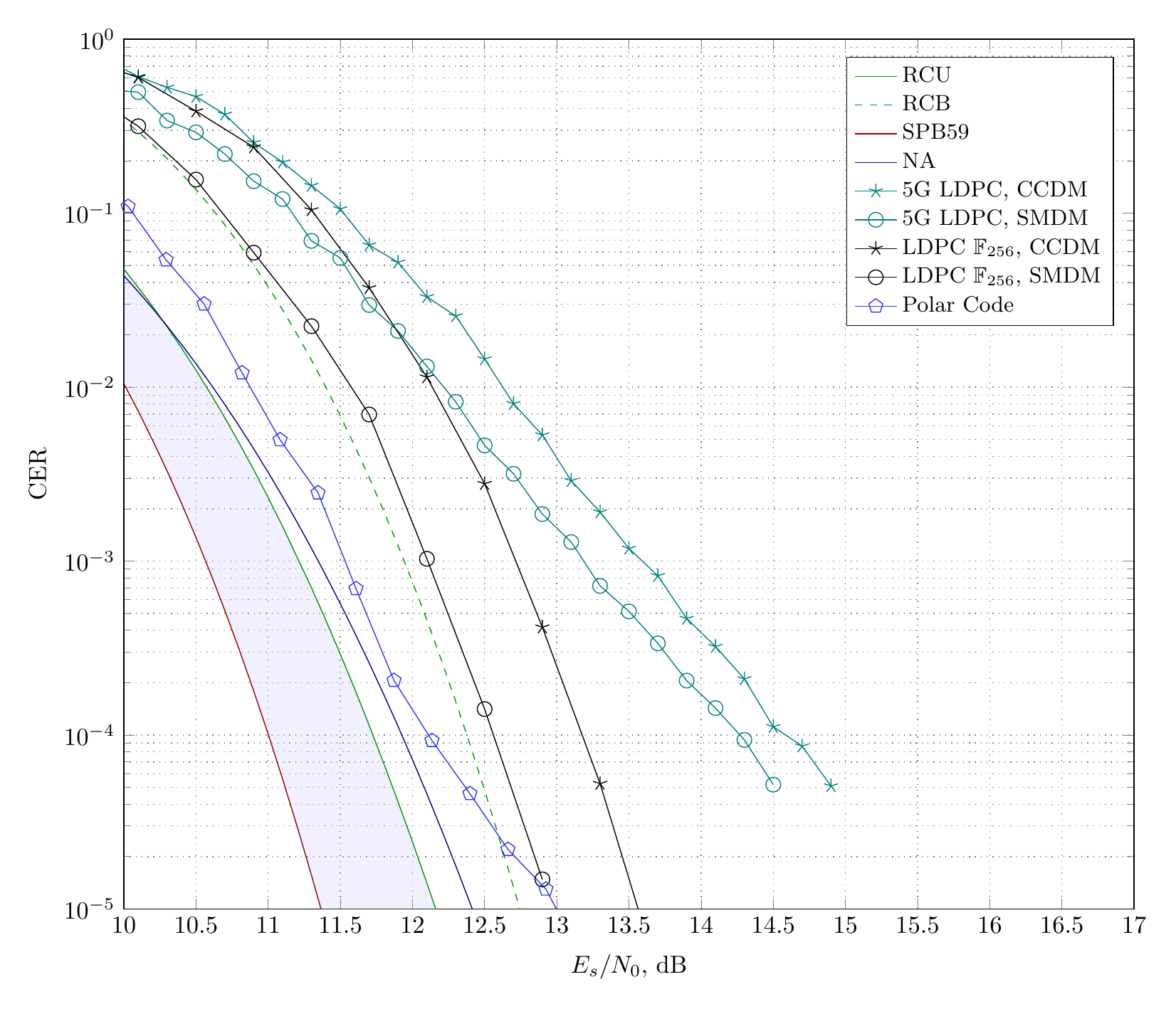}
\caption{Codeword error rate vs. $E_s/N_0$ for $64$-QAM, shaped signaling and an \ac{SE} of \SI{3.0}{bpcu} for several codes over the AWGN channel. The number of channel uses is $32$.}
\label{fig:cmp_192_shaped}
\end{figure}

All coding schemes discussed in the previous sections can be used in a CM scenario with higher-order modulation formats. In Fig.~\ref{fig:cmp_192_uni} and \ref{fig:cmp_192_shaped} we compare CM approaches for a target SE of $3$ bits per channel use  for the case of $64$-QAM and a blocklength of \SI{192}{bits}, i.e., we have a number of $192/6$ = 32 channel uses.
In  Fig.~\ref{fig:cmp_192_uni}, we illustrate a performance comparison for the case of uniform signaling.

\begin{itemize}

\item The binary LDPC code is from the 5G standard~\cite{Richardson5G} and derived from BG~2 (cf. Section~\ref{sec:codesmodern_bin_turbo_ldpc}). We use a random bit-mapper, i.e., the BMD bit channels are assigned randomly to the variable nodes.
\item The NB-LDPC code is an ultra-sparse code of rate $1/2$ and it is constructed over $\ff{64}$. It exhibits a gap of about \SI{0.4}{dB} to the RCB at $\CER=\num{e-4}$.
\item The polar code was designed according to \cite{Boch1703:Efficient} for $E_s/N_0 = \SI{13.45}{dB}$. The list size is $L = 32$ and a 8-bit CRC is used in the concatenation.
\item OSD uses a $(255, 99)$ BCH code that is punctured in 60 parity positions and shortened in $3$ information bit positions to obtain a $(192,96)$ code. The OSD parameter is $t = 4$.

\end{itemize}

In Fig.~\ref{fig:cmp_192_shaped}, we use \ac{PAS} to reduce the shaping loss incurred by uniform signaling and improve the power efficiency. We target a \ac{SE} of $3.0$ bits per channel use, which is achieved by adjusting the DM rate. We show results for two DM approaches, namely \ac{CCDM} and \ac{SMDM}. \ac{CCDM} was proposed first in \cite{schulte_constant_2016} and shown to be the optimal fixed-to-fixed blocklength \ac{DM} for the normalized informational divergence metric and long output blocklengths. Instead, \ac{SMDM} has favorable performance for short blocklengths and is the informational divergence optimal \ac{DM} for finite blocklength. It uses the shell mapping algorithm~\cite{laroia_optimal_1994} internally to perform the mapping to power-efficient channel input sequences.

\begin{itemize}
 \item The binary LDPC code is from the 5G standard~\cite{Richardson5G}, has rate 3/4 and is derived from BG~1 (cf. Sec.~\ref{sec:codesmodern_bin_turbo_ldpc}). We use a random bit-mapper for the amplitude bit levels, the uniform sign bits are assigned to the last variable nodes in the graph. At a CER of $\num{e-3}$ we see that \ac{SMDM} is \SI{0.4}{dB} more power efficient than \ac{CCDM}.
 \item The NB-LDPC code is an ultra sparse cycle code of rate 2/3, constructed over $\ff{256}$. It is operated with \ac{PAS} as discussed in \cite{steiner_ultra-sparse_2017}. The gain of \ac{SMDM} compared to \ac{CCDM} is about \SI{0.6}{dB}.
 \item The polar code was designed according to \cite{Boch1703:Efficient,prinz_polar_2017} for $E_s/N_0 = \SI{12.95}{dB}$. The list size is $L = 32$ and a 4-bit CRC is used  in the concatenation. Additionally, it makes use of a type check during the list decoding to compensate for the finite length losses of \ac{CCDM}~\cite[Sec.~IV]{prinz_polar_2017}.
\end{itemize}

\FloatBarrier

\section{Conclusions}\label{sec:conclusions}

In this paper, we reviewed several code constructions tailored to the transmission of short information blocks. The performance of the codes has been compared with tight information theoretic bounds on the error probability achievable by the best codes.
Our review illustrates that there is a wide spectrum of solutions for efficient transmission in the short-blocklength regime. 
To conclude, we provide a brief list of interesting open directions, which were not addressed in this manuscript:
\begin{itemize}
	\item Some of the decoding  algorithms described in the previous sections are complete, i.e., the  output of the decoder is always a codeword. Incomplete algorithms, such as belief propagation for \ac{LDPC} codes, may output an \emph{erasure}, i.e., the iterative decoder may converge to a decision that is not a (valid) codeword but the error is detected. Hence, while for complete decoders all error events are \emph{undetected}, incomplete decoders provide the additional capability of notifying the receiver when decoding does not succeed. In some applications, it is of paramount importance to deliver very low undetected error rates. This is the case, for instance, for telecommand systems, where wrong command sequences may be harmful. The \ac{CCSDS} \ac{LDPC} code of  Figure \ref{fig:64128} has been designed with this objective in mind, trading part of the coding gain for a strong error detection capability \cite{Dolinar08:AngleMC}. Complete decoders, such as those based on \ac{OSD} and Viterbi decoding, may be used in such critical applications by adding an error detection mechanism. One possibility is to include an outer error detection code. In the short blocklength regime, the overhead incurred by such solution may be unacceptable. In this context, a more appealing solution is provided by a post-decoding threshold test as proposed in \cite{Forney68:error_bounds}. Examples of the application of this approach are given in, e.g., \cite{Hof09:CONV,Hof10:Forney,Williamson14:ROVA}.
	\item The development of codes and decoding algorithms that address channels with unknown state such as fading channels with no \textit{a priori} channel-state information available at the encoder and decoder (see, e.g., \cite{Ostman2018:NC}) is still an open problem. Here, the decoding task is made complicated by the need of accounting for the uncertainty on the channel coefficients. A naive approach is to introduce sufficiently large pilot fields to allow for an accurate channel estimation step. However, when short blocks are transmitted, the use of large pilot fields leads to considerable overheads, i.e., rate losses. This suggests that in this setting channel decoding and channel estimations should be performed jointly (see, e.g., \cite{Coskun:SCC19}).
	\item Throughout the paper, we focused exclusively on the analysis of fixed-length coding schemes. In some applications where communication is bidirectional and a feedback link is hence present, it is more natural to consider variable-length coding schemes with decision (ACK/NACK) feedback. Finite-blocklength bounds for this scenarios are available~\cite{polyanskiy11-08a,ostman18-07a}, but are not as tight as the corresponding bound for the fixed-blocklength case. Also, a more accurate modeling of the ACK/NACK message compared to what is available in the literature may unveil interesting tradeoffs between coding rate and reliability of the feedback information. Indeed, for a fixed frame size, the more channel uses are used for the ACK/NACK message, the less channel uses are available for the coded bits. Code design for this setup have been recently proposed~\cite{williamson15-07a,wang17-06a,yang_joint_2018}. However, the overall code design space is largely unexplored.
\end{itemize} 

\section*{References}
\bibliographystyle{elsarticle-num} 
\biboptions{sort&compress}
\biboptions{square}

\end{document}